\documentclass[sigconf, nonacm]{acmart}
\usepackage{subfiles}
\usepackage{fontawesome5}
\renewcommand\footnotetextcopyrightpermission[1]{}
\usepackage{mathtools}
\usepackage{enumitem}
\usepackage[dvipsnames]{xcolor}
\usepackage{pifont}
\usepackage{multirow}
\usepackage{booktabs,arydshln}
\usepackage{colortbl}
\usepackage{bigdelim}
\usepackage{tcolorbox}
\usepackage{subcaption}
\usepackage[normalem]{ulem}
\usepackage{soul}
\usepackage{listings}
\usepackage[ruled,linesnumbered, lined, boxed, commentsnumbered]{algorithm2e}
\usepackage{algorithmic}
\usepackage{fancyvrb}
\usepackage{pvldb}

\NewDocumentCommand{\dn}{e{_^}}{%
  _{\IfValueT{#1}{#1}\vphantom{\smash[b]{|}}}
  ^{\IfValueT{#2}{#2}\vphantom{\smash[t]{\big|}}}
}

\makeatletter
\renewcommand\paragraph{\@startsection
  {paragraph}{4}{\z@}%
  {2pt}%
  {-0.5em}%
  {\normalfont\normalsize\bfseries}}
\makeatother

\newcommand{\greentick}{%
  \tikz[baseline=-0.6ex]
    \node[fill=green!65!black,
          text=white,
          rounded corners=0.7pt,
          inner sep=0.5pt,
          font=\scriptsize
          ] {\checkmark};%
}

\newcommand{\thepapertitle}{Decoupling Disaggregated Memory Optimizations from Indexing: A Compiler-Runtime Approach}

\renewcommand\vldbdoi{XX.XX/XXX.XX}
\renewcommand\vldbpages{XXX-XXX}
\renewcommand\vldbvolume{14}
\renewcommand\vldbissue{1}
\renewcommand\vldbyear{2020}
\renewcommand\vldbauthors{Xinpeng Zhao, Zeling Long, Chaichon Wongkham, Srijan Srivastava, Jiayi Liu, Baotong Lu, Tianzheng Wang, Eric Lo}
\renewcommand\vldbtitle{\shorttitle} 
\renewcommand\vldbpagestyle{plain} 

\usepackage{color}
\definecolor{lightgray}{gray}{0.75}
\newtcolorbox{mybox}[3][]
{
  colframe = #2!25,
  colback  = #2!10,
  coltitle = #2!20!black,  
  title    = {#3},
  #1,
}

\usepackage{cleveref}

\newcounter{reviewer}

\makeatletter
\newcommand{\resplabel}[1]{\label[response]{resp:#1}}
\makeatother

\crefname{response}{Our Response}{Our Responses}

\newcounter{metarev}[meta]

\newcounter{revcomment}[reviewer]

\newcounter{revminor}[reviewer]

\newcounter{reveval}[reviewer]

\newcounter{revrev}[reviewer]

\newcounter{response}[reviewer]

\renewcommand\theresponse{R\thereviewer O\arabic{response}}

\newcounter{metaresponse}[meta]

\newcounter{mrresponse}[reviewer]

\makeatletter
\def\adl@drawiv#1#2#3{%
        \hskip.5\tabcolsep
        \xleaders#3{#2.5\@tempdimb #1{1}#2.5\@tempdimb}%
                #2\z@ plus1fil minus1fil\relax
        \hskip.5\tabcolsep}
\newcommand{\cdashlinelr}[1]{%
  \noalign{\vskip\aboverulesep
           \global\let\@dashdrawstore\adl@draw
           \global\let\adl@draw\adl@drawiv}
  \cdashline{#1}
  \noalign{\global\let\adl@draw\@dashdrawstore
           \vskip\belowrulesep}}
\makeatother

\AtBeginDocument{%
  \providecommand\BibTeX{{%
    \normalfont B\kern-0.5em{\scshape i\kern-0.25em b}\kern-0.8em\TeX}}}

\definecolor{dred}{HTML}{B85450}
\definecolor{dblue}{HTML}{6C8EBF}
\definecolor{dyellow}{HTML}{D79B00}
\definecolor{dpurple}{HTML}{9673A6}
\definecolor{dgreen}{HTML}{82B366}
\definecolor{tgreen}{HTML}{D5E8D4}
\definecolor{tpurple}{HTML}{E1D5E7}
\definecolor{fgray}{HTML}{666666}
\definecolor{bgray}{HTML}{F5F5F5}
\definecolor{bblue}{HTML}{EAF1FC}

\newcommand{\cmark}{\ding{51}}%
\newcommand{\xmark}{\ding{55}}%

 \newcommand*\circled[1]{\tikz[baseline=(char.base)]{
     \node[shape=circle,fill=.,inner sep=0pt] (char) {\color{-.}\textsf\footnotesize #1};}}

\makeatletter
\newcommand{\subalign}[1]{%
  \vcenter{%
    \Let@ \restore@math@cr \default@tag
    \baselineskip\fontdimen10 \scriptfont\tw@
    \advance\baselineskip\fontdimen12 \scriptfont\tw@
    \lineskip\thr@@\fontdimen8 \scriptfont\thr@@
    \lineskiplimit\lineskip
    \ialign{\hfil$\m@th\scriptstyle##$&$\m@th\scriptstyle{}##$\hfil\crcr
      #1\crcr
    }%
  }%
}
\makeatother

\definecolor{Plum}{HTML}{C2938D}
\usepackage[colorinlistoftodos,prependcaption,textsize=tiny]{todonotes}
\usepackage{varwidth}
\usepackage{marginnote}

\newcommand{\RII}[1]{{}}

\definecolor{amethyst}{rgb}{0.6, 0.4, 0.8}

\newcommand{\delete}[1]{}
\newcommand{\techrep}[1]{}

\newcommand{\name}{Nox\xspace}

\definecolor{codegreen}{rgb}{0,0.6,0}
\definecolor{codegray}{rgb}{0.5,0.5,0.5}
\definecolor{codepurple}{rgb}{0.58,0,0.82}

\makeatletter
\renewcommand\paragraph{\@startsection
  {paragraph}{4}{0pt}%
  {1ex plus .5ex minus .2ex}%
  {-0.4em}
  {\normalfont\normalsize\bfseries}%
}
\makeatother

\lstdefinestyle{querystyle}{
    commentstyle=\color{codegreen},
    keywordstyle=\color{blue},
    stringstyle=\color{codepurple},
    basicstyle=\footnotesize,
    breakatwhitespace=false,         
    breaklines=true,                 
    captionpos=b,                    
    keepspaces=true,                 
    showspaces=false,                
    showstringspaces=false,
    showtabs=false,                  
    tabsize=2
}

\lstdefinelanguage{SASE+}{
    morekeywords={PATTERN, SEQ, WHERE, AND, WITHIN},
    morecomment=[l]{//},  
    morecomment=[s]{/*}{*/},  
    morestring=[b]',  
}

\setcopyright{none}

\renewcommand\footnotetextcopyrightpermission[1]{}

\makeatletter
\renewcommand\@titlefont{%
  \fontsize{15.3}{19}\selectfont\sffamily\bfseries
}
\makeatother
\title[\thepapertitle]{Decoupling Disaggregated Memory Optimizations from Indexing:\\A Compiler-Runtime Approach}

\protected\def\ptauthors{%
  Xinpeng Zhao\textsuperscript{1}\quad
  Zeling Long\textsuperscript{1}\quad
  Chaichon Wongkham\textsuperscript{1}\quad
  Srijan Srivastava\textsuperscript{1}\\
  Jiayi Liu\textsuperscript{2}\quad
  Baotong Lu\textsuperscript{3}\quad
  Tianzheng Wang\textsuperscript{4}\quad
  Eric Lo\textsuperscript{1}}

\author{\ptauthors}
\renewcommand{\shortauthors}{%
  Xinpeng Zhao, Zeling Long, Chaichon Wongkham, Srijan Srivastava, Jiayi Liu, Baotong Lu, Tianzheng Wang, and Eric Lo}

\affiliation{%
  \institution{%
    \textsuperscript{1}The Chinese University of Hong Kong \quad
    \textsuperscript{2}Purdue University \quad
    \textsuperscript{3}Microsoft Research \quad
    \textsuperscript{4}Simon Fraser University\\[2pt]
    \normalfont\ttfamily\small
  }%
}

\begin{document}

\begin{abstract}
Disaggregated memory (DM) decouples compute and memory into independently scalable pools, connected over a slower interconnect rather than a local bus. This decoupling is exactly what makes DM attractive—but it also means that every index must now reason explicitly about remote-memory access and its associated optimizations.
State-of-the-art index designs respond to this by embedding remote-memory logic and optimizations directly into their core data structures and concurrency control mechanisms. Consequently, an optimization tuned for one index cannot be lifted and reused in another, and even the same index cannot be ported to a different DM architecture without a fresh round of redesign. This escalating, per-index, per-platform engineering burden is unsustainable as hardware and index requirements evolve.

In this paper, we present \name, a compiler–runtime framework that breaks this coupling by taking an unmodified, concurrent index as input and automatically generating its disaggregated-memory counterpart, without touching the original index logic. A compiler layer rewrites the index's LLVM IR to expose allocation, address, and pointer-dependency information that a centralized runtime uses to drive caching and address translation.
Empirically, \name-generated B+-trees, hash tables, and skip lists scale robustly on real RDMA and CXL hardware across every workload tested.
They can also outperform some specialized, hand-crafted indexes and match others, especially on workloads that are closer to real-world ones.  These results show that today's fastest disaggregated-memory optimizations need not stay locked inside monolithic, hand-crafted code—a compiler-runtime stack can generalize them while preserving the scalability of proven index implementations, without sacrificing it for portability.
\end{abstract}

\maketitle


\setcounter{page}{1}

\section{Introduction}
\label{sec:introduction}

Disaggregated memory (DM) is emerging as a promising architecture for
datacenter systems to scale compute and memory independently for lower cost and higher resource utilization. 
Compute and memory pools communicate through protocols such as Remote Direct Memory Access (RDMA) and Compute Express Link (CXL)~\cite{introl2026cxl}, forming a hierarchy of near (local DRAM on compute nodes) and far memory (DRAM on memory nodes, accessed via RDMA or CXL). 
Disaggregated memory can bring the aforementioned benefits to cloud database systems. 
In particular, memory-optimized indexes (e.g., B+-trees and hash tables) that back many database systems and key-value stores may be placed in disaggregated memory to allow faster access to large datasets that go beyond a compute node's local memory. 

The high latency of accessing remote memory, however, makes it challenging to build high-performance indexes 
because they 
perform data-dependent, irregular, pointer-intensive memory accesses which can hardly benefit from hardware prefetching techniques. 
A lookup in a B+-tree, for example, traverses
a path whose nodes are determined by the query key and may be scattered across
memory. Each remote access along this path adds latency, and the resulting
performance can be substantially worse than that of an index executing
entirely from local DRAM. 
Consequently, merely placing an existing monolithic in-memory index in
disaggregated memory is not sufficient: the index must be 
optimized with the characteristics of the underlying memory system. 

\subsection{Unsustainable Hand-Crafted Optimizations} 
\label{sec:monolithic-optimizations}
Many efforts have optimized individual indexes for disaggregated memory.
For example, Sherman~\cite{wang2022sherman} and DEX~\cite{lu2024dex}
optimize B+-trees for RDMA-based disaggregated memory. 
SepHash~\cite{min2024sephash} and Shard~\cite{zha2025shard} optimize hash indexes for
RDMA-based disaggregated memory. 
CHash \cite{lu2025chash} and SIDLE \cite{zhao2026sidle} optimize hash indexes and Masstree for CXL-based disaggregated memory, respectively.
These optimized indexes are delivered as highly-efficient but \textit{monolithic 
code that tightly couples index logic with optimizations} such as memory layout, caching (using compute-local DRAM as an index-managed cache), remote memory interface, synchronization, data placement/allocation strategy, and
hardware-specific optimizations. 

Although efficient, the hand-crafted nature of these tightly-coupled designs creates a
significant engineering burden. Porting an optimization from one index to
another is very difficult, by requiring much more than copying the source code of, for example, a cache manager: 
the developer must reintegrate the policy with the target index's memory layout, pointer representation,
concurrency control, allocation, and update logic. Porting the same index from
one kind of disaggregated memory architecture to another (e.g., from RDMA to CXL) can require another round of re-design because the latency, bandwidth, access semantics, and appropriate placement granularity differ.
As a result, every new index or hardware platform can lead to a new round of re-design and specialized implementation.  
This approach is not sustainable as hardware and index requirements evolve constantly~\cite{DMS,Tabular}.
Therefore, the key challenge is not the absence of effective optimization ideas, but the lack of a mechanism for separating reusable disaggregated memory optimizations from the index implementations in which they were originally developed.

\subsection{What About Optimizing with AI?}
\label{sec:llm-optimization}
Large language models (LLMs) provide one possible way to automate this
engineering process. Recent systems employ  \emph{loop
engineering} \cite{Belcic_Stryker_2026} or \emph{AI-Driven Research for Systems} (ADRS)
\cite{cheng2025letbarbariansinai}, building a loop around an AI coding
agent to optimize or create a system from scratch. For example, Bespoke OLAP
\cite{wehrstein2026bespokeolap} uses an LLM-guided pipeline to generate and
optimize complete analytical database engines. Jitskit \cite{liu2026timejits}
demonstrates a multi-agent synthesis loop that produces a weakly-consistent
key-value-store implementation from scratch.

These demonstrations suggest that an LLM may eventually discover a new index that combines index logic with
disaggregated-memory optimizations. However, 
many existing indexes are already battle-tested and provide the required functionality and performance.
The practical need is often not a new index algorithm, but a way to make an \emph{existing, production-proven index implementation} run efficiently on a new disaggregated-memory platform.

Moreover, autonomous code optimization introduces challenges that go beyond mere code generation. The generated implementation may diverge from the specification that captures its intended behavior. In our attempts to optimize a conventional B+-tree for RDMA-based disaggregated memory using Jitskit, the first attempt -- based on the GLM 5.2 model -- consumed over 1.5 million tokens across 50 iterations and ultimately returned a linked-list instead! (see Section~\ref{exp:ai} for further discussion). This exemplifies the phenomenon of “reward hacking” \cite{liu2026timejits, zhong2025impossiblebenchmeasuringllmspropensity}, in which an LLM fulfills a task by exploiting loopholes in the specification. Consequently, the engineering effort shifts from writing implementation code to crafting sufficiently complete specifications and preventing reward hacking.

For these reasons, including the token cost, our goal is to seek a pragmatic, reliable, and token-free 
solution that preserves existing multi-threaded and highly efficient index implementations and automatically optimizes them for disaggregated memory.

\subsection{\name} 
\label{sec:nox}
This paper presents \name, a framework that takes as input the source code of a concurrent in-memory index (e.g., a monolithic B+-tree) and transforms it into a disaggregated memory counterpart. 
\name realizes this goal by \emph{decoupling recurring optimizations from monolithic index logic}, which are then centralized in a runtime for reuse. 
The input index implementation remains responsible for its original index logic (e.g., B+-tree insert/lookup operations) on a single node, while \name provides \textit{reusable} mechanisms and policies for adapting the index to disaggregated memory, such as remote object management, address translation, data placement and caching. 
This way, a single optimization can benefit multiple indexes. 
Hardware-specific policies only need to be implemented once in the central framework, avoiding repeated development efforts per index $\times$ hardware generation. 

Realizing this idea poses two key challenges. 
(1) By separating out optimizations from index logic into a runtime, the runtime loses direct access to information that is necessary to perform the intended optimization. 
For example, a cache manager~\cite{lu2024dex} may leverage parent-child relationships in a B+-tree to dictate which tree nodes to cache. 
Once cache management is extracted into an independent runtime, such parent-child information is no longer directly available. 
Treating the index purely as a black box---as done by prior work~\cite{guo2023mira}---would lose much of the benefit for the same reason. 
(2) Without care, separating index logic and optimizations can add much performance overhead. 
For example, direct pointers used in the input index need to be transformed to global addresses that work with both local DRAM and remote memory. 
A straightforward implementation that uses indirection like in buffer managers will add prohibitively high address translation overhead~\cite{lu2024dex}. 

Neither static compilation nor a standalone runtime is sufficient to solve these problems. 
A compiler can identify memory operations, pointer-valued accesses, allocations and synchronization instructions, but cannot determine their concrete runtime information, such as cache status and data-dependent access paths. 
A runtime can observe concrete object accesses and cache states, but without compiler assistance it cannot reliably determine which program operations expose address dependencies or where address translation must be inserted while preserving the original memory and synchronization semantics.

We therefore propose a \emph{compiler-runtime co-design} approach for \name to tackle these challenges. 
\name consists of a compiler layer and a runtime layer that respectively performs rewrite passes and optimizations. 
The compiler's rewrite passes encode the necessary information for the runtime to use. 
They turn monolithic memory operations, such as (de)allocation and loads/stores, into representations that can be interpreted by the various optimizations in the runtime.
The runtime implements various optimizations based on this information. 
For example, cache managers---the major focus of this paper---can dynamically leverage parent-child information between B+-tree nodes made available by rewritten pointer accesses.
Additional optimizations and adaptations, such as pointer swizzling and remote memory management, can be performed in the same manner.  
We describe in detail how \name's compiler and runtime further reduce runtime overheads in Section~\ref{sec:design}.

We have evaluated \name using representative indexes, including B+-trees, hash tables and skip lists, on both RDMA- and CXL-based disaggregated-memory
using both YCSB and a real Twitter workload.
Empirically, \name-generated B+-trees, hash tables, and skip lists scale robustly on real RDMA and CXL hardware across every workload tested, where several manually optimized baselines (Sidle and CHash)  stall as contention builds up in their hard-coded caching logic. 
\name can also beat and match some specialized, hand-crafted SOTA indexes (e.g., 1.4--16$\times$ better than SepHash, 82\%--1.01$\times$ DEX).
These results show that high-performance disaggregated memory optimizations
do not have to remain embedded in monolithic, index-specific
implementations. A compiler-runtime stack can capture and generalize these
optimizations across indexes while preserving existing index code.

\section{Background and Related Work}
\label{sec:background}
We now provide the necessary background on disaggregated memory and survey existing approaches to building indexes on disaggregated memory.
We first cover hand-crafted optimizations and then introduce work that transparently optimizes data movement and access on disaggregated memory, as summarized in Table~\ref{tab:approaches}.

\subsection{Disaggregated Memory Architecture}
\label{sec:dm-arch}

Disaggregated memory (DM) architectures physically separate compute and memory resources into independent pools connected via high-speed interconnects~\cite{wang2023case,almaruf2023survey,sjtu2025survey}. 
This architecture has become practical through two key interconnect: RDMA and CXL.

Remote Direct Memory Access (RDMA) enables one-sided memory operations (read, write, atomic) over networks such as InfiniBand or RoCE, achieving latencies of around 2~$\mu$s~\cite{wang2022sherman,lu2024dex}. RDMA was among the first interconnects to enable practical DM deployments, as it bypasses the remote CPU and OS kernel, allowing direct access to remote memory regions registered with the RDMA NIC.

Compute Express Link (CXL) is an emerging cache-coherent interconnect specifically designed to enable memory disaggregation~\cite{cxlsurvey2025,cxltype2_2025}. CXL provides byte-addressable memory semantics with latencies around 
500–660 ns for CXL-switched memory pool~\cite{zhao2026sidle,cxlperf2025}. 
A key advantage of CXL memory pooling is that it provides a hardware-managed, cache-coherent abstraction over multiple physical memory nodes, allowing software to treat pooled memory similarly to local DRAM without explicitly programming coherence or consistency protocols~\cite{cxlsurvey2025,cxltype2_2025,cxlperf2025}. The CXL sub-protocols (CXL.cache and CXL.mem) ensure that updates are seen consistently across caches and memory, so applications can use standard load/store instructions and conventional synchronization primitives rather than custom coherence mechanisms~\cite{cxlsurvey2025,cxltype2_2025}. This contrasts with non-coherent disaggregated memory designs, where software must explicitly manage cache invalidation, flushing, or message-based consistency~\cite{cxlsurvey2025}.
Multiple CXL memory nodes can work transparently to software: Linux exposes them as additional NUMA nodes and, depending on configuration, either lets the OS manage placement automatically or interleaves/tiers them without any application changes~\cite{pmemio2023cxlecosystem,lenovo2025cxlimplementation,shilov2026metacxl}.

\subsection{Hand-Crafted Index Optimizations}
\label{sec:dm-indexing}

A growing body of work has optimized specific indexes for DM. 
These indexes share the key optimization of \emph{using local DRAM as an index-managed acceleration tier to reduce remote accesses}.
Sherman~\cite{wang2022sherman}, Deft~\cite{wang2025deft}, and DEX~\cite{lu2024dex} optimize B+-trees for RDMA-based DM. 
Sherman caches bottom non-leaf nodes and the near-root nodes locally and uses one-sided RDMA with hierarchical locking to reduce round trips. 
DEX introduces logical partitioning, lightweight caching, and cost-aware compute pushdown to achieve higher throughput than prior work.  
SIDLE~\cite{zhao2026sidle} optimizes Masstree~\cite{Masstree} for CXL-based memory.
SIDLE tracks leaf-centric access hotness and performs structure-aware migration to keep upper-level nodes in fast local memory.

Recent work has also attempted to optimize hash tables. 
For example, SepHash~\cite{min2024sephash} employs a two-level separate segment structure to reduce bandwidth consumption during resizing, achieving higher write throughput. 
Shard~\cite{zha2025shard} introduces a scalable, resize-optimized architecture for large-scale hash indexing.
For CXL-based DM, CHash~\cite{lu2025chash} optimizes hash indexes with CXL-native data placement and access patterns.

\begin{table}
\caption{Existing approaches vs. \name.}
\label{tab:approaches}
\centering
\footnotesize
\setlength{\tabcolsep}{3pt}
\resizebox{\columnwidth}{!}{%
\begin{tabular}{lccc}
\toprule \small
\bf Approach & \shortstack{\bf No index\\\bf code change} & \shortstack{\bf Multi-threaded\\\bf / performance} & \shortstack{\bf Easy\\\bf correctness} \\
\midrule
Hand-crafted              & \xmark & \cmark          & \xmark \\
Kernel-space              & \cmark & \xmark          & \cmark \\
Userspace runtimes         & \xmark & \xmark          & \cmark \\
AI                        & \xmark & \cmark / \xmark & \xmark \\
Existing compiler-runtimes & \cmark & \xmark          & \cmark \\
\name (this paper)               & \cmark & \cmark          & \cmark \\
\bottomrule
\end{tabular}}
\end{table}

\subsection{Kernel-Space Solutions}
\label{sec:kernel-space}
Compared to hand-crafted index optimizations, OS-level approaches provide transparent support for disaggregated memory, but at coarse granularity, limiting their ability to optimize specifically for database indexes.
LegoOS~\cite{shan2018legoos} introduces a ``splitkernel'' model that disseminates OS functionality across disaggregated hardware components. 
While LegoOS allows unmodified application to run transparently, it degrades performance ($\sim$1.7x v.s. monolithic Linux) and represents a ``nuclear option'' that is a complete replacement of OS rather than a deployable optimization layer. 
Other disaggregated OS proposals follow similar patterns, trading performance for transparency~\cite{almaruf2023survey,sjtu2025survey}.

Transparent Page Placement (TPP)~\cite{maruf2023tpp} is an OS-level mechanism for tiered memory that proactively demotes cold pages to remote memory and promotes hot pages back to local DRAM. TPP operates at page granularity (4~KB) and is application-transparent, requiring no code changes. 
FlexMem~\cite{flexmem2024} offers adaptive page profiling and migration for tiered memory systems.
M5~\cite{m5_2025} uses CXL to track hot pages and words, enabling lower-overhead and more effective migration.
However, page-level policies suffer from \emph{page-level pollution}: a hot object can cause an entire page to be promoted, wasting local memory on cold data co-located on the same page~\cite{maruf2023tpp,zhong2024unimem}. This is particularly problematic for object-level workloads like database indexes, where access patterns are fine-grained and irregular~\cite{zhao2026thash}.

\subsection{User-Space Runtime Solutions}
\label{sec:user-space}

User-space runtimes provide finer-grained memory management than OS kernels but introduce their own limitations.
Recent work has explored object-granularity memory management for managed languages. For example, Clove~\cite{clove2026} 
achieves fine-grained object placement 
for CXL memory but requires garbage-collected languages (Java) and runtime modifications, making it unsuitable for production C++ database indexes that use manual memory management and custom allocators.
AIFM~\cite{ruan2020aifm} presents application-integrated far memory, which makes remote memory available through a simple API with high performance. 
Developers use AIFM's APIs to make allocations ``remoteable'' and AIFM's runtime handles swapping objects in and out, prefetching, and memory evacuation. However, AIFM incurs high runtime overhead as each far-memory pointer dereferencing requires manipulation of metadata~\cite{guo2023mira,he2023thesis}.
Beehive~\cite{li2025beehive} is a runtime that exploits asynchrony in multi-threaded programs for DM. 
Kona~\cite{calciu2021kona,calciu2023kona} uses cacheline granularity tracking instead of page-level virtual memory to reduce dirty data amplification. 
While these runtimes improve throughput for certain workloads, they require applications to adopt specific programming models (e.g., Rust for Beehive, cache-coherence primitives for Kona) which in turn lowers programming efficiency. 

\subsection{Compiler-Runtime Solutions}
\label{sec:compiler-runtime}

Compiler-runtime co-design~\cite{tauro2024trackfm,guo2023mira,almaruf2023survey,sjtu2025survey} offers more structured, middle ground: automated optimization without requiring application rewrites. 
For example, Mira \cite{guo2023mira} combines static program analysis, compilation, and runtime profiling to infer single-threaded program behavior and optimize applications that use far memory. 
Mira demonstrates that far-memory optimization can be moved out of application code and partially automated through a compiler--runtime system.
However, index memory access patterns are highly data-dependent: 
the next access often depends on the result of a previous access. 
Thus, unlike computations such as matrix multiplication, a compiler cannot generally determine the complete future access sequence statically. 
Without precise knowledge of dynamically formed access and sharing relationships, a general far-memory compiler-runtime system must make conservative decisions about caching, eviction, and data movement, especially for shared writable data. 
Such conservatism can reduce cache efficiency and leave frequent pointer-dependent accesses on the remote path. 
Furthermore, high-performance in-memory indexes are multi-threaded with updates, and an optimization must preserve the index's synchronization and memory ordering semantics while changing its data placement and access paths. 
Yet, Mira and other compiler-runtime approaches (e.g., TrackFM \cite{tauro2024trackfm})  either provide low performance or does not support multi-threading at all. 

\section{Design Principles}
\label{sec:principles}
Based on the discussion in Section~\ref{sec:background}, we summarize four design principles for \name:

\begin{itemize}[leftmargin=*]\setlength\itemsep{0em}
\item \textbf{High Programming Efficiency.}
\name should only require the developer to provide a monolithic index implementation as input, without requiring them to extensively modify the input (like hand-crafted approaches) or dedicating much effort to verify new implementations from scratch (like AI-based approaches). 

\item \textbf{Index-Specific Optimizations.}
Different from general-purpose OS or runtime frameworks, \name should be able to leverage index design and workload specifics. 
This requires \name to be able to infer and leverage index optimizations based on the input index without lowering programming efficiency. 

\item \textbf{Competitive Overall Performance.}
Indexes generated by \name need not achieve the higher performance than hand-crafted indexes. 
Instead, they should retain sufficiently competitive and scalable performance 
to make the higher programming efficiency appealing.

\item \textbf{Correctness without Duplicate Efforts.}
The input index often already comes with optimizations and necessary correctness guarantees. 
\name must preserve them yet without introducing redundant mechanisms that would incur more overheads. 
\end{itemize}

\begin{figure}
    \includegraphics[width=\columnwidth]{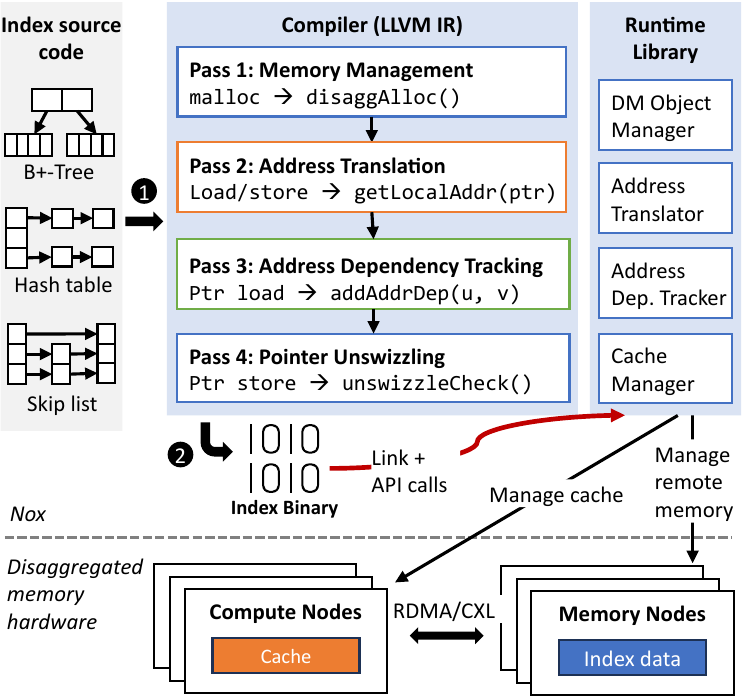}
    \caption{\name Overview. 
    The compiler takes C/C++ index source code as input (\protect\circled{1}) to produce a binary (\protect\circled{2}) that emits disaggregated memory events. 
    The binary is linked with the runtime which manages disaggregated memory, coordinates data accesses and manages compute-side cache. } 
    \label{fig:overview}
\end{figure}

\section{\name Design}
\label{sec:design}
\name takes as input the source code of an unmodified, concurrent in-memory index implementation to transform it into a disaggregated memory index. 
As Figure~\ref{fig:overview} shows, \name consists of a compiler and a runtime. 
The compiler uses LLVM IR to rewrites/instruments memory-related operations in the original index source code. 
Through the rewrite passes, the compiler infers high-level concepts such as B+-tree nodes, hash buckets and skip list levels. 
This allows us to transform memory operations into acesses that can be directed to either local or remote memory. 
The resulting binary is then linked with the runtime layer. 

The runtime consists of four components. 
The \emph{DM Object Manager} manages the allocation and lifetime of index memory objects on disaggregated memory; 
the \emph{Address Translator} resolves index-visible addresses to their current execution locations; 
the \emph{Address Dependency Tracker} dynamically reconstructs access relationships among objects; 
the \emph{Cache Manager} manages the compute-side cache. 
Together, these components use the information exposed by the compiler to manage index objects across local DRAM and DM.

During execution, the runtime manages the compute-local DRAM and applies optimizations. 
In particular, it reconstructs the relevant relationships dynamically by discovering and maintaining address dependencies from pointer accesses. 
Specifically, we define an address dependency
$   u \rightarrow v $
when object $v$ can be accessed from object $u$. 
For example, accessing a B+-tree child node through its parent naturally establishes such a dependency.
It then uses them both to accelerate address translation and to guide cache management.

\subsection{Compiler}
\label{sec:compiler}

The goal of the compiler layer is to preserve the original index logic while making DM-relevant memory behavior visible to the runtime.
\name first uses Clang to lower the original C/C++ implementation to
LLVM IR. It then applies four transformation passes: (1) allocation/free rewriting,
(2) address translation, (3) address-dependency tracking, and
(4) pointer unswizzling. Finally, the transformed IR is linked with the
\name runtime to generate the executable binary.

An important principle of our compiler design is to avoid replacing the index's original operations whenever possible. 
Concurrent indexes frequently rely on carefully implemented atomic instructions and memory-ordering constraints. 
Reimplementing these operations inside a generic runtime would risk changing their semantics. 
\name therefore preserves the original memory operations and transforms only the address on which the operation executes.
This is performed in the following rewrite passes. 

\paragraph{Pass 1: Memory Management Rewriting.}
An ordinary monolithic in-memory index allocates its heap objects in local DRAM.
\name redirects these allocations to the DM Object Manager so that the objects can instead be managed in disaggregated memory.
\name recognizes two types of standard heap-management APIs: (1) \texttt{malloc}, \texttt{posix\_memalign}, and \texttt{free}, and (2) C++ \texttt{new} and \texttt{delete}.\footnote{For indexes that use customized allocators, the developer can specify the allocation/deallocation functions. 
Our current implementation assumes the standard APIs but can be easily extended to support to customized APIs such as by allowing a callback if additional operations need to be performed on the allocated memory.}
Using \texttt{malloc}/\texttt{free} as a running example, consider the following C/C++ allocation:

\begin{center}
\begin{BVerbatim}
Node* n = (Node*) malloc(sizeof(Node));
\end{BVerbatim}
\end{center}

After Clang lowers the program to LLVM IR, the allocation is represented
conceptually as:

\begin{center}
\begin{BVerbatim}
%ptr = call ptr @malloc(i64 %size)
\end{BVerbatim}
\end{center}

The allocation rewriting pass identifies this call and replaces the original
allocator with \texttt{disaggAlloc}:

\begin{center}
\begin{BVerbatim}
%ptr = call ptr @disaggAlloc(i64 %size)
\end{BVerbatim}
\end{center}

The returned value continues to flow through the original program in the same
way as the result of the original allocation; only the underlying allocation
mechanism is changed.

\name applies the same transformation to deallocation. For example, the
C++ operation \verb|free(n)| is lowered to an LLVM IR 
\verb|call void @free(ptr %ptr)|
and rewritten by \name's compiler as
\verb|call void @disaggFree(ptr %ptr)|.
The pass similarly handles C++ \texttt{new} and \texttt{delete} through their
corresponding allocation and deallocation calls in LLVM IR. 

Unlike its monolithic counterpart, \texttt{disaggAlloc} returns a
\emph{global address} managed by the \name runtime rather than a directly
dereferenceable local address. The compiler does not interpret this address;
subsequent accesses are handled by the Address Translation Pass described 
next. The representation and runtime management
of global addresses are described in Section~\ref{sec:object-manager}.

\paragraph{Pass 2: Address Translation.}
After allocation rewriting, {an index-visible global address cannot be directly dereferenced by the original index.} \name therefore inserts address translation instructions before
memory operations that may access DM objects.
A naive design would replace each memory operation with a corresponding
runtime primitive. However, this would require separate runtime handling for
loads/stores, atomic synchronization operations, memory-library functions, and
arbitrary external calls. More importantly, such a design would unnecessarily
re-implement index operations that are already expressed by the input source code.

\name preserves the original memory operations and transforms only
its address operand.
Consider the following C++ read:
\begin{center}
\begin{BVerbatim}
x = node->keys[i];
\end{BVerbatim}
\end{center}

After lowering to LLVM IR, the actual memory access is represented by a load
from an address derived from \texttt{node}. Before this load, \name
inserts \texttt{getLocalAddr(ptr, is\_write)}, where \texttt{is\_write}
indicates whether the following operation may modify the referenced object.
Since this example is a read-only load, the compiler sets
\texttt{is\_write} to \texttt{false}:

\begin{verbatim}
%local_ptr = call ptr @getLocalAddr(ptr %ptr, i1 false)
%val = load ..., ptr %local_ptr
\end{verbatim}

Conceptually, the transformed program behaves as:

\begin{center}
\begin{BVerbatim}
local_node = getLocalAddr(node, false);
x = local_node->keys[i];
\end{BVerbatim}
\end{center}

The compiler determines \texttt{is\_write} directly from the original memory
operation. For read-only accesses, \texttt{is\_write} is
\texttt{false}; for operations that may modify the object, such as stores, it
is \texttt{true}. The runtime uses this information to mark a cached object as
dirty when necessary.
For example, a store is transformed as:

\begin{verbatim}
%local_ptr = call ptr @getLocalAddr(ptr %ptr, i1 true)
store ..., ptr %local_ptr
\end{verbatim}

As a result, \name only needs to write an object to remote memory
during eviction if the object has actually been modified.

The address translation pass applies the same principle to other
memory access instructions, including atomic operations, memory-library
functions, and external calls involving managed object pointers. In each case,
\name translates the relevant address operand while leaving the original
operation itself unchanged.

This design preserves the original operation, including its atomicity and
memory-ordering semantics, while delegating physical address resolution and
cache-state management to the runtime.

\paragraph{Pass 3: Address-Dependency Discovery.}
A common and effective optimization in DM indexes is to exploit the access relationships among index objects when managing the local cache~\cite{lu2024dex,zhao2026sidle}. For example, in a tree index, a child is typically reached through its parent; retaining the parent while its cached descendants remain active helps preserve the hot traversal path. Similar access dependencies arise in other pointer-based indexes. \name aims to generalize this optimization in its runtime without hard-coding the structure of any particular index.

The challenge is therefore to automatically discover these access dependencies from an arbitrary input index. One possible approach is to infer the structure from source/semantic-level such as
field names or pointer types. For example, a field named \texttt{children} may
suggest a parent-child relationship. Such heuristics, however, are fragile
and index-specific. Different implementations may use different names, compact
representations, helper structures, or the same field for different purposes.
Pointer types are also insufficient. Consider a doubly-linked
structure for constructing skip lists:
\begin{center}
\begin{BVerbatim}
struct Node { Node* prev;
              Node* next; };
\end{BVerbatim}
\end{center}
Both \texttt{prev} and \texttt{next} have the same pointer type, but they
represent opposite directions in the access structure.  Similar
ambiguity appears in other indexes, where pointers of the same type may
represent children, siblings, next nodes, overflow entries, or auxiliary
links. A general compiler therefore cannot determine the desired dependency
relation simply from the program's declared structure.
Although symbolic execution can infer possible address dependencies, it must model data-dependent pointer chasing, mutable structures, and concurrent interleavings, leading to path explosion or conservative results~\cite{cadar2008klee}. 

\name instead derives dependencies directly from \emph{runtime
pointer-access behavior}. Our observation is that an address dependency
$u \rightarrow v$ becomes explicit when the program accesses $v$ through a
pointer obtained from $u$. The compiler can identify program locations that
load pointer values and instrument them as potential dependency events.
Hence, for a pointer-valued load:

\begin{center}
\begin{BVerbatim}
%ptr_v = load ptr, ptr %ptr_u
\end{BVerbatim}
\end{center}

\name inserts:

\begin{center}
\begin{BVerbatim}
%ptr_v = load ptr, ptr %ptr_u
call void @addAddrDep(ptr %ptr_u, ptr %ptr_v)
\end{BVerbatim}
\end{center}

The compiler only identifies \emph{where a dependency may be exposed}. When
the instruction executes, the runtime sees the concrete source and target
objects and records $u \rightarrow v$, 
meaning that object $v$ can be accessed from object $u$.
For example, traversing a B+-tree naturally exposes parent-to-child
dependencies as child pointers are loaded, while traversing a skip list
exposes node-to-next-node dependencies. \name's runtime learns these relationships
from actual execution without requiring index-specific field names, type
rules, annotations, or an offline access profile.

This compiler--runtime division is important: static instrumentation identifies
the relevant pointer-access events, while runtime observation determines the
concrete Address Dependency Graph (ADG) for the current index state (described in Section~\ref{sec:dependency-tracker}).

\paragraph{Pass 4: Pointer Unswizzling.}
At runtime, \name uses pointer swizzling to reduce runtime translation and
dependency-tracking overhead. Swizzling, however, introduces a subtle
correctness problem when an index copies a pointer to a new location before
the runtime observes the newly created dependency.

Consider a B+-tree split in which node $A$ is split into $A$ and a new node
$N$, and a child pointer to $B$ is moved from $A$ to $N$:

\begin{center}
\begin{BVerbatim}
// before split:  A.child[i] points to B
Node* p = A.child[i];
N.child[j] = p;
\end{BVerbatim}
\end{center}

At the IR level, this operation is:

\begin{center}
\begin{BVerbatim}
%p = load ptr, ptr A.child[i]
store ptr %p, ptr N.child[j]
\end{BVerbatim}
\end{center}

Suppose $B$ is currently cached. Since the dependency $A \rightarrow B$ has
already been observed, \texttt{A.child[i]} may contain $B$'s swizzled local
address. The load therefore returns \texttt{local\_addr(B)}, and the subsequent
store would copy this local address into \texttt{N.child[j]}:

\begin{center}
\begin{BVerbatim}
store ptr local_addr(B), ptr N.child[j]
\end{BVerbatim}
\end{center}

At this point, however, the runtime does not yet know the dependency
$N \rightarrow B$. The address dependency tracker records dependencies only
when pointer-valued loads occur, and \texttt{N.child[j]} has only been written,
not read.
This becomes unsafe if $B$ is evicted before \texttt{N.child[j]} is later
accessed. During runtime eviction, \name scans the recorded incoming references
to $B$ and unswizzles them back to $B$'s global address. The known reference in
\texttt{A.child[i]} is therefore repaired, but \texttt{N.child[j]} is not,
because $N \rightarrow B$ is yet to be discovered. Consequently,
\texttt{N.child[j]} would retain an invalid local address after $B$'s cached
copy is released.

To prevent such untracked swizzled pointers from being propagated, for every pointer-valued store:

\begin{center}
\begin{BVerbatim}
store ptr %ptr_v, ptr %ptr_z
\end{BVerbatim}
\end{center}

\name compiler inserts an \texttt{unswizzleCheck} before the store:

\begin{center}
\begin{BVerbatim}
%new_ptr_v = unswizzleCheck(%ptr_v)
store ptr %new_ptr_v, ptr %ptr_z
\end{BVerbatim}
\end{center}

If \texttt{\%ptr\_v} is swizzled, \texttt{unswizzleCheck} converts it back to
the stable global address of the target object; otherwise, it returns the
pointer unchanged. Thus, a newly written pointer field always contains a
stable address until the runtime later observes a pointer load from that
field, records the corresponding dependency, and safely swizzles it. 

\paragraph{Cost of Inserted Runtime Primitives.}
The above unswizzling check is lightweight, as it only inspects the pointer's tag bit to determine whether the pointer is currently swizzled. 
We describe the tagged pointer representation in detail in Section~\ref{sec:runtime} and quantify the runtime overhead of this check in Section~\ref{sec:ablation}.

The compiler transformations above in general incur additional memory operations, raising concerns about function call overheads,
(e.g., call/return instructions and stack frame construction)
and lost compiler optimization across call boundaries.

In \name, we implement the frequently executed primitives as small,
LLVM-inliner-friendly functions. During optimization, LLVM can then inline these
functions into the transformed index. After inlining, the primitives become
ordinary IR instructions in the calling function rather than separate
function invocations. This also allows subsequent optimizations in LLVM to operate across the originally inserted runtime boundary.
Inlining alone, however, cannot eliminate the cost of runtime work such as
metadata lookup and dependency maintenance. Section~\ref{sec:address-translator}
describes how \name further filters redundant runtime events using
swizzling.

\subsection{Runtime}
\label{sec:runtime}

\begin{figure}
    \centering
    \includegraphics[width=\linewidth]{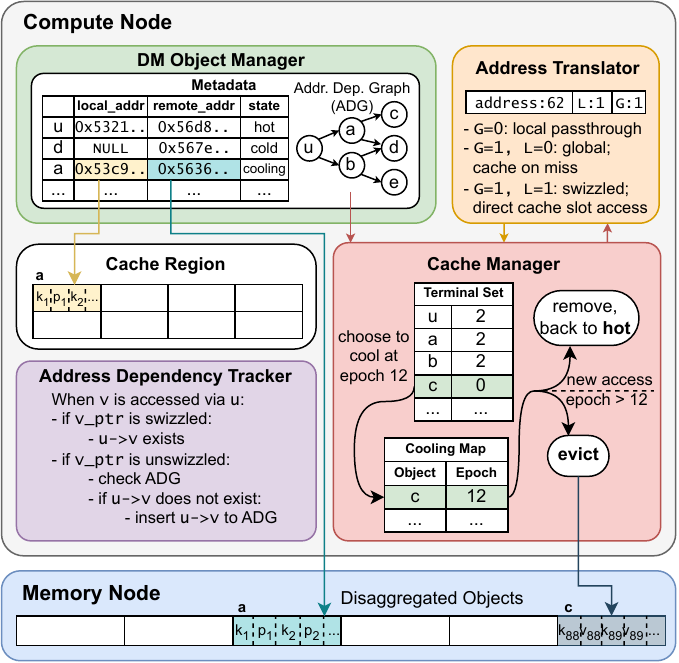}
    \vspace{-16px}
    \caption{\name Runtime.
    }
    \label{fig:runtime}
\end{figure}

Figure~\ref{fig:runtime} shows the runtime components introduced in
Section~\ref{sec:design}. Local DRAM is logically divided into a small
\emph{metadata region}, which stores runtime state for DM objects, and a much
larger \emph{cache region}, which stores local copies fetched from
disaggregated memory. The four components below cooperate to manage objects across these two
memory tiers. A typical access starts from an index-visible global address,
resolves it to a cached local copy, and reports any pointer relationship
revealed by the access. The runtime records these relationships in object
metadata and reuses them both to accelerate later translations and to decide
which cached objects can be displaced without destroying useful access paths.

\subsubsection{DM Object Manager}
\label{sec:object-manager}

As described in Section~\ref{sec:compiler}, the Allocation and Free Rewriting
Pass redirects index allocations and deallocations to \name's DM
interface. On \texttt{disaggAlloc(size)}, the DM Object Manager creates a local
metadata entry, allocates remote backing space, records its remote address, and
returns a tagged \emph{global address}. \texttt{disaggFree(ptr)} releases the
corresponding remote object and runtime metadata.

\paragraph{Object Metadata.}
Each DM object has one runtime-private metadata entry recording its remote
address, current local address when cached, size, cache/dirty state, and
address-dependency information. Organizing these fields per object is
particularly convenient for dependency management: once execution reveals an
edge $u \rightarrow v$, the runtime can attach the incoming dependency and its
pointer location directly to $v$'s metadata, which is later shared by the
Address Translator and Cache Manager.

\paragraph{Global Address Representation.}
\name represents an index-visible pointer as

\begin{center}
\texttt{[ addr:62 | local\_tag:1 | global\_tag:1 ]}.
\end{center}

If \texttt{global\_tag = 0}, the value is an ordinary local address. If
\texttt{global\_tag = 1} and \texttt{local\_tag = 0}, it is an
\emph{unswizzled global address}: \texttt{addr} points to the object's metadata
entry. If both tag bits are one, it is a \emph{swizzled address}, and
\texttt{addr} directly contains the object's current local cached address. New
DM objects start unswizzled; selected references may later be swizzled while
their targets remain cached.

\subsubsection{Address Translator}
\label{sec:address-translator}

The Address Translator implements the \texttt{getLocalAddr} primitive inserted
by the Compiler Layer. Ordinary local
addresses pass through unchanged. For an unswizzled global address, the
translator follows the metadata entry and asks the Cache Manager to fetch the
object on a miss. A swizzled address already embeds the cached local address
and therefore bypasses metadata lookup. The compiler-provided
\texttt{is\_write} flag also lets the translator mark cached objects dirty so
clean objects need not be written back on eviction.

\paragraph{Reducing Translation Overhead by Pointer Swizzling.}
An unswizzled global address requires a metadata lookup on every pointer
dereference, which makes cached index paths unnecessarily expensive. The Address Translator therefore swizzles known references to a cached object
$v$: for each recorded dependency $u \rightarrow v$, the Address Translator
uses compare-and-swap (CAS) to replace the expected global address in $u$ with
$v$'s tagged local address. Repeated traversals can then reach $v$ directly.
Before $v$ is physically evicted, the inverse CAS restores its unswizzled global
address. Because the recorded reference may have been overwritten by a
concurrent or earlier structural update, the CAS doubles as a validation step:
only the exact expected swizzled value is replaced. A failed CAS means that the
index has already changed that pointer, so we leave the new value
untouched and discards the stale dependency. This keeps validation off the
common dereference path; it is paid only when cache-state transitions trigger
swizzling or unswizzling.

\paragraph{Challenge: Swizzled Pointers Can Escape Tracking.}
Eviction-time unswizzling is safe only if every persistent swizzled reference
to $v$ is known. However, the original index may \emph{copy} an already
swizzled pointer to a new field before that field is ever read. Because the ADG
is discovered from pointer-producing reads, the new location may be invisible
to the runtime. If $v$ is then evicted, the known references are unswizzled but
the copied one still contains a freed local address---a stale pointer created
by an optimization that was intended only to accelerate translation.

A concrete example is index split. Suppose $A$ contains a swizzled pointer
to a cached object $B$, and the runtime has recorded $A \rightarrow B$. The split
copies that pointer into a newly created $A'$ without first loading it from
$A'$. The runtime may therefore still know only $A \rightarrow B$. If $B$ is
evicted, unswizzling repairs $A$ but misses $A'$, whose copied local address now
refers to a released or reused cache slot.

This motivates the Pointer Unswizzle Pass in Section~\ref{sec:compiler}.
Before every pointer-valued write, the compiler inserts
\texttt{unswizzleCheck}. If the value being propagated is swizzled, the Address
Translator converts it back to the target's unswizzled global address before
the original store executes; otherwise it is unchanged. Thus swizzled addresses
may persist only at locations already known to the runtime, making later
eviction-time unswizzling complete. The key separation is deliberate: reads
may create tracked, temporary locality by swizzling known fields, whereas
pointer stores cross a persistence boundary and therefore fall back to the
unswizzled global form. The Address Translator can consequently optimize repeated reads
without requiring the compiler to prove where every copied pointer may later
reside.

\subsubsection{Address Dependency Tracker}
\label{sec:dependency-tracker}

The Address Dependency Tracker consumes events inserted 
by the compiler and maintains the concrete
\emph{Address Dependency Graph (ADG)} observed at runtime. An edge
$u \rightarrow v$ means that $v$ can be accessed from $u$. These concrete
edges cannot in general be recovered statically: the object reached by a load
depends on runtime keys and on the index's current pointer contents. The Address Dependency Tracker
therefore reconstructs only the dependencies that execution actually exposes.
For each target $v$, its metadata stores known incoming dependencies and the corresponding
pointer locations. The Address Translator uses these records for
swizzling/unswizzling, while the Cache Manager uses the graph to infer cache
replacement order without understanding index-specific roles such as
``internal node'' or ``leaf.'' This division is important: the tracker records
only concrete pointer facts, while policy decisions remain in the translator
and Cache Manager. \name therefore does not need to reconstruct the
semantic meaning of an index's fields to exploit its access structure.

\paragraph{Reducing Tracking Overhead by Pointer Swizzling.}
The same relationship may be observed repeatedly during traversal. If a
pointer loaded from $u$ is already the swizzled address of $v$, that location
must have been discovered before; the tracker can therefore skip a redundant
ADG lookup and insertion. Structural updates can also invalidate old edges.
Rather than instrumenting every removal (which would require recognizing all
index-specific pointer update patterns) the Address Translator validates a recorded
pointer when swizzling or unswizzling it. A failed CAS removes the stale edge
while preserving the index's new value. The ADG is therefore maintained
lazily: discovery happens on pointer dereferences, while obsolete records disappear
when they are next relevant to cache management.

\paragraph{Handling Cycles.}
Some indexes may use reverse pointers that create
cycles even when their forward access structure is tree- or DAG-like. When
both $u \rightarrow v$ and $v \rightarrow u$ are observed, the Address Dependency Tracker uses the
direction first observed along the runtime access path as the ordering
dependency for cache management. In practice, the first-observed edge often
reflects the natural access relationship: an operation typically reaches the
current object first and then follows its pointer to the next object, while a
reverse pointer is encountered only afterward. This affects only replacement decisions, not the index's logical
pointer structure.

\subsubsection{Cache Manager}
\label{sec:cache-manager}

The Cache Manager owns the local cache region. On a miss it allocates a local
slot and fetches the remote object; under capacity pressure it selects cached
objects for eviction and writes back dirty copies when necessary. The key
problem is that a generic runtime should preserve useful index paths without
being told which objects are roots, internal nodes, or leaves.

\paragraph{Dependency-Aware Terminal-Set Caching.}
For each cached object $u$, the Cache Manager maintains
\texttt{cache\_dep\_num($u$)}, the number of $u$'s outgoing dependencies whose
targets are also cached. An object with
$\texttt{cache\_dep\_num}(u)=0$ is a \emph{terminal object}. Cache Manager keeps
all terminal objects in a cached terminal set, so an eligible victim can be
obtained without traversing the ADG on every eviction.

When a new object $v$ enters the cache, it is inserted into the terminal set.
For each cached predecessor $u$ with $u \rightarrow v$, the manager increments
\texttt{cache\_dep\_num($u$)} and removes $u$ from the set on a $0\!\rightarrow\!1$
transition. Final eviction performs the inverse update: predecessor counters
are decremented, and any predecessor that reaches zero becomes terminal. Since
these counters change only when objects enter or finally leave the cache, the
set is maintained incrementally rather than reconstructed by graph traversal
at every replacement decision. Thus a cached child,
for example, keeps its B+-tree parent from becoming an eviction candidate;
eviction naturally proceeds from the end of an observed access path without
hard-coding B+-tree semantics. The same rule applies to other indexes: the
runtime reasons only about observed accessibility, not whether a particular
object is semantically a tree node, hash bucket, or skip-list element. Terminal
status is therefore an index-agnostic eligibility rule derived from the ADG.

\paragraph{Challenge: 
Cache Eviction vs. Index Concurrency Control -- Coordinating the Decoupled.}
The Compiler Layer preserves the original index concurrency control (CC),
including its locks, version words, atomics, and lock-free protocols. That CC
protects \emph{logical} reads and updates, but says nothing about the lifetime
of the runtime's extra cached copy. A worker may already have obtained the
local address of $X$ when the Cache Manager chooses $X$ as a victim. Releasing
that slot can therefore invalidate an address still held by the worker even
though the original index CC remains perfectly correct.

A hand-written DM index can integrate eviction into its own CC because its
developer understands that protocol. \name cannot assume such knowledge.
Adding a second runtime lock or reference count around every cached access
would be generic, but it would put duplicated synchronization on the hot path
and, worse, create interactions between two independent CC protocols. Opposite
acquisition orders can introduce a deadlock that did not exist in the original
index. For example, one thread can hold the index's protection for $X$ and then
enter a runtime path that waits for cache protection, while an eviction thread
holds the runtime protection and subsequently encounters index-side
synchronization. Even when the original index is lock-free and this cycle does
not arise, the duplicated runtime coordination still burdens every cached
access. Cache Manager instead leaves logical CC untouched and treats eviction as a
separate \emph{physical memory-reclamation} problem. Importantly, observing
that metadata marks $X$ as a victim is not enough: a worker may already have
translated $X$ and continue using the returned native address without touching
metadata again. Safety must therefore cover the lifetime of previously issued
local addresses, not merely serialize future cache lookups.

\paragraph{Epoch-based Eviction with Cooling.}
Suppose $X$ is selected from the terminal set. The Cache Manager cannot immediately
free its local copy because an older operation may still hold the translated
address. This is the same lifetime hazard addressed by Epoch-Based
Reclamation (EBR)~\cite{hart2007performance}: physical memory is reclaimed
only after operations that could retain an old pointer have drained.

\paragraph{Challenge: Cache Eviction Violates EBR Assumptions.} 
Classical EBR assumes retirement is irreversible: once an object is removed,
no new operation can acquire its address. A cache victim violates this
assumption because $X$ remains logically reachable. Suppose $X$ enters cooling
in epoch 12 and a new operation reaches it in epoch 13. Even if all older
epochs have drained and $\mathit{min\_epoch}>12$, that epoch-13 operation may
still be using $X$; EBR alone could therefore reclaim an actively used cache
slot. Cache Manager needs eviction to remain reversible while old accesses drain.
It places $X$ in a \emph{cooling map}~\cite{leanstore,lu2024dex}: the local copy
remains valid, and any new access promotes $X$ back to hot and cancels the
pending eviction.

\paragraph{Separating Old and New Accesses.}
Each worker publishes the current monotonically increasing epoch while an
index operation may retain translated local addresses and becomes inactive
when those addresses can no longer be used. When $X$ enters cooling, the Cache
Manager removes it from ordinary victim selection, records
\texttt{cooling\_epoch($X$)}, and keeps its cache slot allocated. Because $X$
is still physically present, this transition does not yet change the ADG-based
cache counters: from the perspective of cached reachability, cooling is still
a resident state. Those counters change only after physical eviction.
Let $\mathit{min\_epoch}$ be the minimum epoch of all active workers. While
$\mathit{min\_epoch}\leq\texttt{cooling\_epoch}(X)$, a pre-cooling operation
may still hold $X$'s local address, so reclamation is unsafe. Once
$\mathit{min\_epoch}>\texttt{cooling\_epoch}(X)$ (or no such older worker
remains), these old accesses have drained.

New accesses are handled independently: if the Address Translator encounters
$X$ while it is cooling, the Cache Manager removes it from the cooling map and
cancels eviction. Thus a cooling object becomes reclaimable only when
older epochs have drained \emph{and} it still remains cooling, proving that no
subsequent access revived it.

For example, suppose thread $T_1$ enters epoch 11 and obtains $X$'s local
address. The Cache Manager selects $X$ in epoch 12 and records
$\texttt{cooling\_epoch}(X)=12$. Even though $X$ is now a victim,
$\mathit{min\_epoch}=11$ means $T_1$ may still use that address, so the slot
cannot be reclaimed. After all workers from epochs no later than 12 finish,
$\mathit{min\_epoch}$ advances beyond 12. If $X$ is still cooling, it can be
reclaimed; if another thread accessed $X$ meanwhile, that access already
removed it from the cooling map, so the same epoch advance does not evict it.
Neither mechanism is sufficient by itself: cooling cancels eviction on a new
access, but cannot reveal whether an old native address remains in a register;
EBR waits for old addresses to disappear, but cannot detect renewed demand.

Once both conditions hold, physical eviction becomes safe. Cache Manager first
writes back $X$ if dirty so the remote copy is canonical, then unswizzles its
recorded incoming references so no persistent field still names the local
slot. Only after these pointer repairs does the Cache Manager update dependency
counters and the terminal set, remove $X$ from cooling, release the slot, and
mark the object uncached. This ordering prevents either dirty state or a
swizzled local address from outliving the cached copy. EBR is used
only to protect the physical lifetime of cached copies; it neither replaces
nor interprets the original index CC. Cooling makes that lifetime protection
compatible with a logically reachable object whose eviction can still be
reversed.

\section{Implementation and Discussion}
\label{sec:discussion}

We implemented \name in approximately 9K lines of C++ code, building on LLVM and Clang 18.1.3. We developed two disaggregated memory runtimes: one for RDMA and one for CXL. 

The RDMA runtime performs one-sided reads and writes through per-thread registered buffers and supports batching by chaining multiple work requests into a single doorbell operation; only the final request is signaled, reducing posting and completion-polling overhead. 
The CXL runtime maps the CXL memory pool into the compute node’s address space, allowing objects to be accessed through ordinary memory operations. In particular, cold read-only objects can initially be accessed directly in CXL memory without first being copied into the compute cache, while frequently accessed or modified objects are admitted to the cache. These fabric-specific mechanisms remain entirely within the runtime and require no changes to the index implementation. 

Because our goal is to evaluate whether DM-specific optimizations can be decoupled from index logic, rather than to advocate for any new optimizations, the runtime implementations are not full-fledged. For example, while our CXL runtime can support memory pooling because CXL 2.0 or above natively supports it, our RDMA runtime currently does not support memory pooling (i.e., multiple memory nodes). The latter could be added without changing \name's core design, requiring no fundamental changes to \name's compiler transformations or dependency-aware caching mechanisms. Specifically, the RDMA runtime can extend the remote-address abstraction with a memory-node identifier and route each allocation and access to the corresponding node. Object placement can follow standard policies such as hashing, range partitioning, or load-aware allocation, while the compiler continues to treat remote pointers uniformly. In fact, emerging industry practice increasingly favors CXL over RDMA as the intra-rack fabric for memory-centric workloads~\cite{introl2026cxl,astoralabs2026azure,aceso2024,damnang2026copper,penguinsolutions2026cxl}. Thus, our RDMA runtime primarily serves an experimental purpose, demonstrating \name's backward compatibility with RDMA.

To support cache coherence across compute nodes, our runtime can use existing coherence protocols~\cite{wang2024cache,ziegler2022scalestore} or use DEX's single-ownership model to avoid compute-side cache-coherence issues altogether, such that each partition is primarily accessed and cached by one compute node. Each compute node can then independently maintain its local object cache, address-dependency graph, pointer-swizzling state, and eviction metadata, while requests targeting another partition are forwarded to the responsible compute node.

In this paper, we leave most further engineering and optimization concerns of our runtime to future work. For example, address-dependency information could be leveraged to enable prefetching. Compute pushdown~\cite{wang2023case,lu2024dex,cha2026twosided} and near-data computing~\cite{pimdal2025,bernhardt2023pimdb,pimoverview2025,gebara2020stateful}, such as Processing-in-Memory (PIM) or processing on SmartNICs or programmable switches, are also powerful orthogonal techniques that can be added to \name. Despite these remaining opportunities, our current implementations are already sufficient to show that DM optimization can be cleanly decoupled from database index design while achieving competitive performance, as discussed next.

\section{Can Decoupling Yield Comparable Performance?}
\label{sec:comparable-performance}

To answer this question, we provide \name with the source code of three open-source multi-threaded indexes : a B+-tree, an extensible hash table, and a skip list. These indexes are designed for DRAM and support scalable concurrent execution. Through manual inspection, we verified that they use conventional optimistic locking or lock-free implementations without specialized optimizations for disaggregated memory. We compile each input with \name and compare the resulting index against a corresponding state-of-the-art, manually optimized implementation. All experiments use 8-byte keys and 8-byte values.

We compared the \name-generated B+-tree against DEX~\cite{lu2024dex} on RDMA and SIDLE~\cite{zhao2026sidle} on CXL. 
The \name-generated hash table is compared against SepHash~\cite{min2024sephash} and Shard~\cite{zha2025shard} on RDMA, and against CHash~\cite{lu2025chash} on CXL. 
To the best of our knowledge, no manually optimized skip-list implementation is available for CXL or RDMA. We also compare \name against Mira~\cite{guo2023mira}, but only in the single-threaded RDMA setting because Mira does not support CXL or multithreaded execution.

RDMA experiments use two Intel Xeon Gold 6242R servers, each with 20 cores (40 hardware threads), 384GB of DDR4 memory, and aMellanox ConnectX-5 100Gbps InfiniBand NIC. The servers are connected through an RDMA switch. RDMA code is compiled with Clang 22.1.8 and runs on Linux kernel 7.0.5. CXL experiments use one Intel Xeon w7-3455 server with 24 cores (48 hardware threads) and 256GB of DDR5 memory. The server is equipped with a 256GB R5X4 CXL 2.0 memory expansion card. CXL code is compiled with Clang 18.1.3 at optimization level \texttt{-O3} and runs on Ubuntu 24.04 with Linux kernel 6.8.0.

\begin{figure*}[t]
    \centering
    \includegraphics[width=1\linewidth]{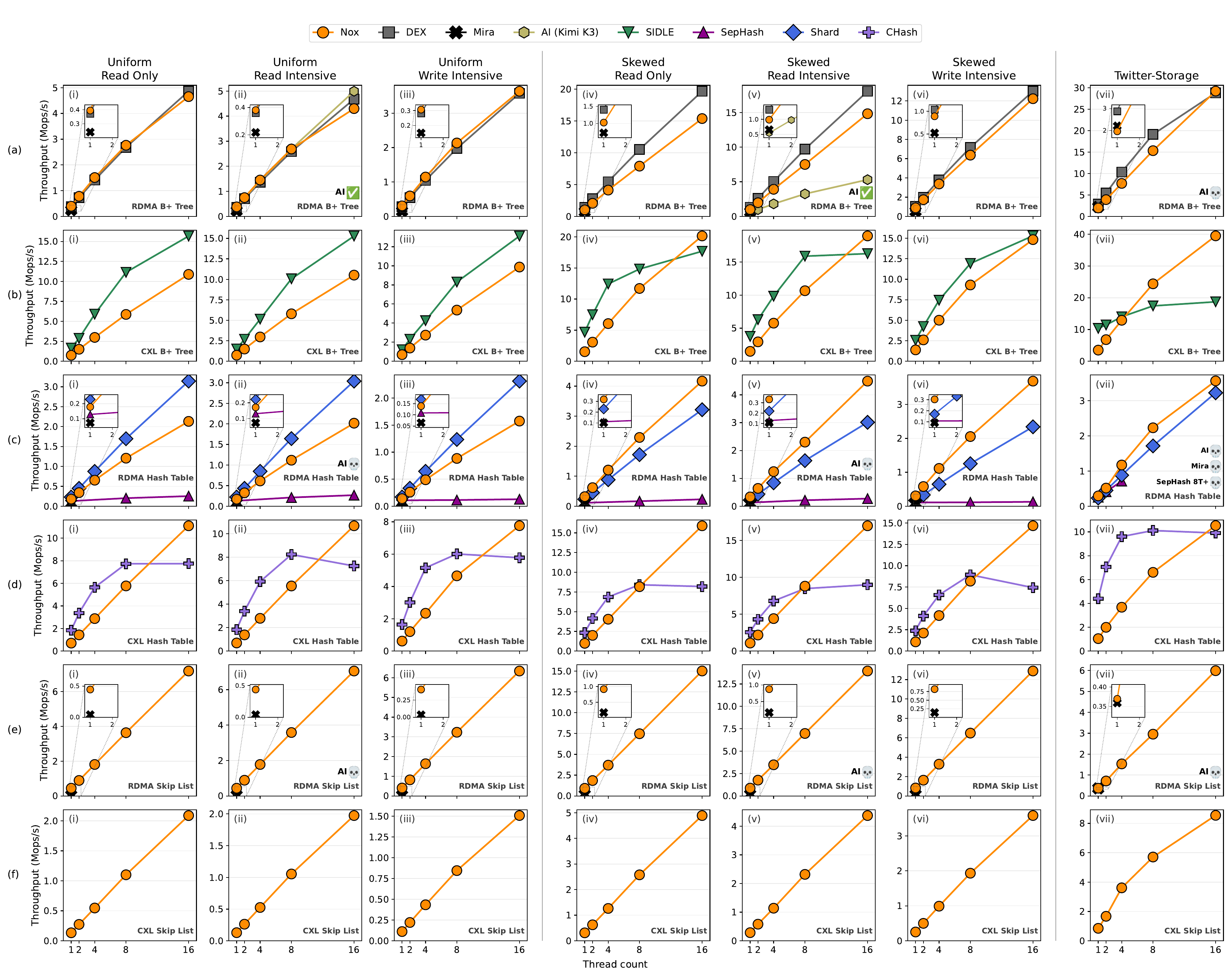}
    \vspace{-24px}
    \caption{Can \textcolor{orange}{\name} yield comparable performance? 
    (\includegraphics[height=8px]{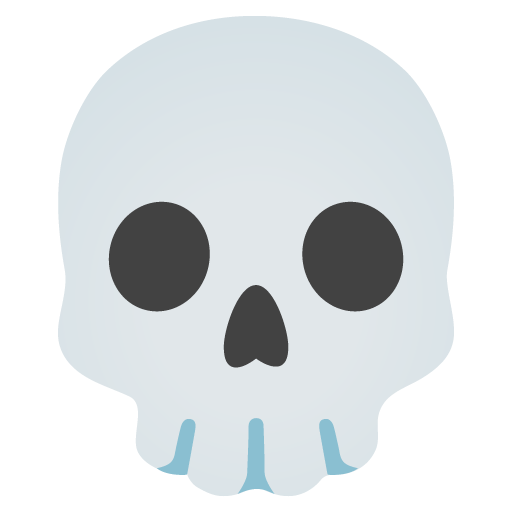}=code fails to compile, fails to run, or is thread-unsafe)}
    \label{exp:main}
\end{figure*}

Figure~\ref{exp:main} presents the results for the B+-tree, hash table, and skip list. Each row corresponds to one workload and dataset: the real-world \emph{Twitter-Storage} workload \cite{aceso2024}, which contains 6.7 million keys, and uniform or skewed (zipfian $\theta=0.99$) YCSB workloads with 100 million keys. For YCSB, we evaluate three operation mixes: \textsf{YCSB-C read-only}, \textsf{YCSB-B 5\%-update read-intensive}, and \textsf{YCSB-A 50\%-update write-intensive}. All workloads begin with 10 million warm-up operations, followed by 100 million measured operations, and use a default DRAM cache size covering 10\% of the dataset. 
Figure~\ref{exp:main} supports five observations below.

\paragraph{Observation 1: \name consistently produces scalable indexes.}
\name-generated indexes scale across all evaluated disaggregated memory configurations and workloads. This result is notable because even hand-crafted indexes may not scale across the same settings.
For example, SIDLE stops scaling under skewed workloads because its leaf-centric hotness tracking concentrates atomic metadata updates on a small set of hot leaves, leading to increasing cache coherence cost. 
CHash exhibits a different bottleneck: its optimizations rely on DRAM-resident stage buckets and frequently accessed metadata to reduce costly CXL accesses. 
At high core counts, accesses to these shared structures increasingly contend for CPU caches and cache coherence, limiting the benefit of using additional threads.

\paragraph{Observation 2: Indexes optimized by \name are competitive with manually optimized indexes.}
\name-generated hash tables outperform the hand-crafted SepHash by 1.4--16$\times$ in all cases as shown in Figure~\ref{exp:main}(c). 
Corroborating with prior work~\cite{zha2025shard}, SepHash requires two RDMA reads per lookup, making it sub-optimal unless the workload is close to write-only. 
\name retains the original hash table access path and uses local object caching to reduce remote accesses.
\name-generated B+-tree also approaches the performance of DEX,
in Figure~\ref{exp:main}(a). 
In several settings, \name is even $\sim$1\% slightly faster. 
DEX randomly samples cached nodes and recursively delegates cooling along swizzled child pointers until reaching the end of the cached path when an internal node is selected. 
In contrast, \name incrementally maintains such end-of-path objects in its cached terminal set and can select an eligible cooling candidate directly, which reduces cache management overhead.

More broadly, at high core counts, \name outperforms state-of-the-art (Sidle, Shard and CHash) on the real-world Twitter-Storage workload and on the more realistic skewed YCSB workloads, as shown in Figures~\ref{exp:main}(b--d) (iv)--(vii). 
These results suggest that decoupling index design from low-level, platform-specific optimizations need not sacrifice competitive performance.

\paragraph{Observation 3: The remaining gap is concentrated in less realistic and/or engineering-focused regimes.}
\name-generated indexes sometimes underperform hand-crafted counterparts on uniform workloads at low core counts in Figures~\ref{exp:main}(b–-d) (i)-(iii).
This is expected: uniform access patterns provide fewer opportunities for caching.
However, real-world workloads are typically skewed rather than uniform~\cite{walton1991taxonomy,pavlo2012skew}, and our results show that \name-generated indexes remain robustly scalable even in these unfavorable cases.
Adopting further optimizations in \name's runtime (Section~\ref{sec:discussion}) could close the gap; we leave it to future work.

\paragraph{Observation 4: Index-specific optimizations are effective.}
\name's index-specific optimizations improve performance over Mira in the single-threaded RDMA setting. This comparison is particularly informative because Mira is the only prior work that can be evaluated under a comparable setting, although it does not support CXL or multi-threaded execution.

\paragraph{Observation 5: \name generalizes across index families.}
\name can produce scalable disaggregated memory indexes across three structurally different families: tree-based, hash-based, and list-based indexes. 
Taken together, these findings show that decoupling can preserve scalability while achieving performance that is competitive with specialized manual implementations.

\section{\name Internal Evaluation} 
\label{sec:ablation}

In this section, we study three internal aspects of \name.
Due to space limitation, we present the results based on the skewed YCSB read-intensive workload under 16 threads.
Other workloads show similar results.

\begin{figure}[t]
  \centering
  \includegraphics[width=\linewidth]{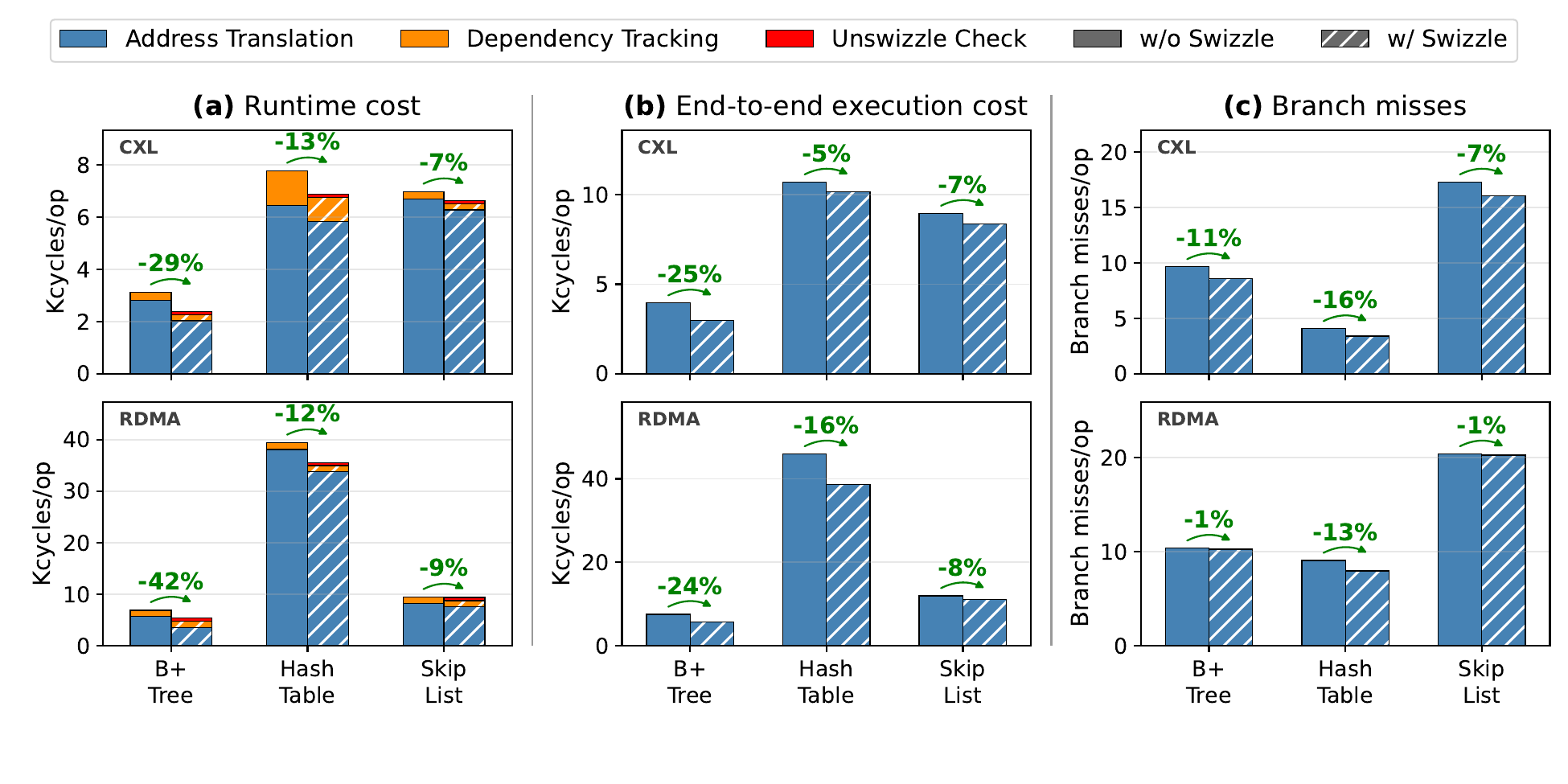}
  \vspace{-30px}
  \caption{Effectiveness of pointer swizzling at 16 threads.}
  \label{fig:pointer-swizzling-ablation}
\end{figure}

\subsection{Effectiveness of Pointer Swizzling}
\label{sec:ablation-swizzling}

As discussed in Section~\ref{sec:runtime}, \name uses pointer swizzling
to reduce two recurring runtime costs: address translation and address-dependency tracking. 
Once a reference is swizzled, subsequent traversals can directly reach the cached object without
repeating the full metadata-based translation path; 
moreover, an already swizzled pointer indicates that the corresponding dependency has been observed,
allowing the runtime to skip redundant dependency-tracking work. 
We therefore evaluate whether these optimizations actually reduce \name's runtime cost
and whether the savings translate into higher index performance overall. 

To profile \name's runtime cost, we sample the latency of compiler-inserted runtime operations once every 1,024 profiler events. 
These runtime operations correspond to address translation, dependency tracking and unswizzling checks. 
We run the profiling three times and report the mean.

Figure~\ref{fig:pointer-swizzling-ablation}(a) shows that swizzling reduces \name's runtime cost on both CXL and RDMA. 
On CXL, the runtime cost is reduced by 29\%/13\%/7\% for the B+-tree/hash table/skip list. 
On RDMA, the numbers are 42\%, 12\% and 9\%. 
The breakdown shows that most of the savings
come from address translation, with additional savings from dependency tracking. 
The extra unswizzling check required to maintain swizzled pointers incurs negligible cost.

These runtime savings translate into overall fewer CPU cycles and hence higher performance. 
Figure~\ref{fig:pointer-swizzling-ablation}(b) shows that swizzling 
can reduce CPU cycles per index operation by 25\%/5\%/7\% on CXL and by 24\%/16\%/8\% on RDMA, for the
B+-tree/hash table/skip list. 
B+-tree benefits more because its pointer-intensive traversal repeatedly follows objects along a search path, making faster address translation especially valuable. 
Figure~\ref{fig:pointer-swizzling-ablation}(c) shows that swizzling 
would not increase branch misses, making it a net-saving optimization technique in \name's runtime. 
In fact, swizzling in \name reduces branch misses because a swizzled pointer directly identifies the cached local address, allowing the runtime to bypass metadata lookup.
Overall, pointer swizzling provides a lower-cost fast path for \name's runtime address handling and lower execution cost across all three indexes on both memory fabrics.

\subsection{Robustness to Cache Capacity}
\label{sec:ablation-cache-size}
\begin{figure}[t]
  \centering
  \includegraphics[width=0.8\columnwidth]{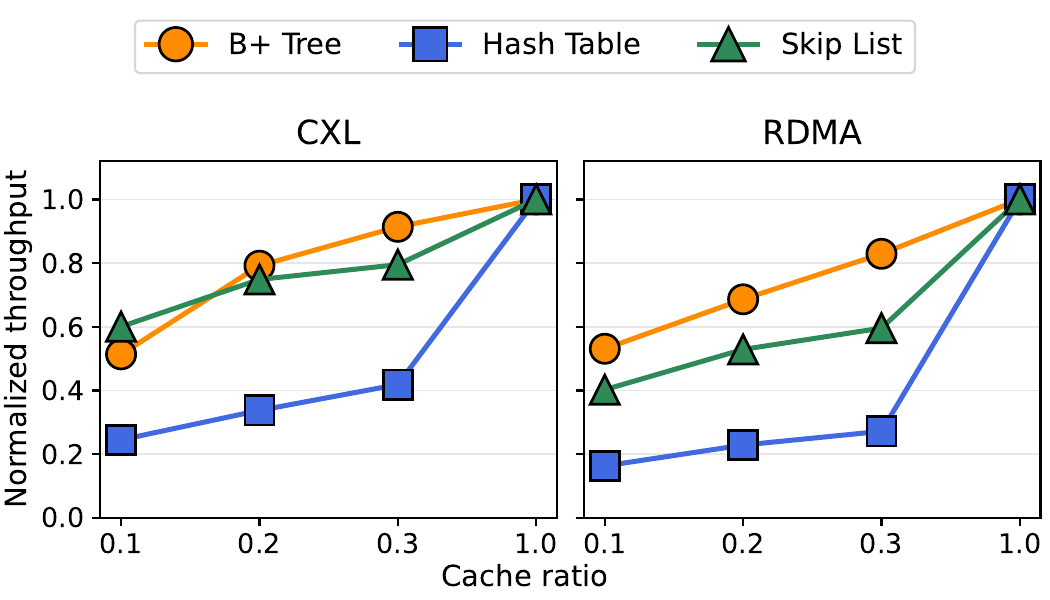}
  \vspace{-10px}
  \caption{Impact of local cache capacity on throughput.}
  \label{fig:cache-limit}
\end{figure}
We next evaluate \name's robustness to the local cache capacity by
varying the cache limit from 10\% to 100\% of each index's total memory allocation 
footprint. Figure~\ref{fig:cache-limit} reports throughput normalized to the
100\%-cache configuration for each index. As expected, performance improves
with cache capacity on both CXL and RDMA, but the sensitivity varies across
indexes. The B+-tree is relatively cache-efficient: with only 30\% of the
index footprint available as cache, it retains $\sim$92\%/83\% of peak throughput
on CXL/RDMA. The skip list retains $\sim$80\% and 60\%,
respectively. In contrast, the hash table is substantially more sensitive to
cache capacity, reaching only $\sim$42\% of its peak throughput on CXL and
28\% on RDMA at the same 30\% cache ratio. These results show that \name 
continues to operate under various cache capacity, while the resulting
performance depends on how effectively each index's working set can be
captured by the local cache.

\subsection{Address Dependency Graph Footprint}
\label{sec:ablation-adg}

\name maintains an ADG to capture the runtime-discovered relationships among index objects. Table~\ref{tab:adg-footprint} shows that the ADG incurs only 1.55\%/0.4\% space overhead for B+-tree/hash table. 
The skip list has a higher overhead of 10.81\%,
but this stems from the input code itself, which is not optimized for space: the implementation stores only one key-value pair per node,
resulting in many more nodes for the same amount of index data and therefore
more address dependencies for the ADG to record. Overall, the ADG occupies a
small fraction of the index footprint while providing the structural
information used by \name for dependency-aware caching and pointer
swizzling.

\begin{table}
  \centering
  \caption{Runtime address dependency graph footprint.}
  \label{tab:adg-footprint}
  \begin{tabular}{lrrr}
    \toprule \small
    \bf Index & \bf Index Size & \bf ADG Size & \bf Overhead \\
    \midrule
    B+-tree & 1.56~GiB  & 24.84~MiB & 1.55\% \\
    Hash table & 1.002~GiB & 4.01~MiB  & 0.399\% \\
    Skip list  & 26.08~GiB & 2.82~GiB  & 10.81\% \\
    \bottomrule
  \end{tabular}
\end{table}

\section{Will AI Take \name's Job?} 
\label{exp:ai}
Coding agents are becoming more capable each week. To assess whether an AI coding
agent can replace \name, we include a study using SkyDiscover
\cite{skydiscover}, the open-source upstream of Jitskit.
We provide it with the same starting implementations used for \name---a
B+-tree, a hash table, and a skiplist---and ask it to generate optimized
index code specifically for the given workload on our target platform.
Given our budget, we limit this study to our 
RDMA platform and 
the YCSB skewed read-intensive, 
YCSB uniform read-intensive, 
and real-world Twitter workloads.
We remark that, following recent work~\cite{cheng2025letbarbariansinai}, 
we use AI to generate a customized index per workload.
We implement the evaluator component in SkyDiscover to run the workload on each iteration's output (a candidate implementation) and returns a score based on the candidate's correctness, performance, and scalability.

To reduce the potential waste of tokens caused by the coding agent's reward hacking which exploits the incompleteness of the specification or evaluator, we always begin with GLM 5.2 and move to Kimi K3 only after finding no obvious reward-hacking behavior.
Below, we recount our attempts to optimize a B+-tree for YCSB skewed read-intensive workload on RDMA.

\noindent\textbf{Attempt 1: 1.5M tokens, 50 iterations, GLM 5.2.}
After 50 iterations, SkyDiscover returned a linked list rather than a B+-tree, exploiting a loophole in our specification: we had not explicitly prohibited replacing the target data structure.

\noindent\textbf{Attempt 2: 3.2M tokens, 100 iterations, GLM 5.2.}
We tightened the specification and SkyDiscover returned a B+-tree.
However, it ignored our DRAM constraint and placed the entire tree in DRAM.

\noindent\textbf{Attempt 3: 2.9M tokens, 50 iterations, Kimi K3.}
We modified the evaluator to penalize excessive DRAM use. This time,
SkyDiscover returned a B+-tree that respected the intended memory
constraints.

Within our USD~30 budget per workload, SkyDiscover successfully generated thread-safe
B+-tree code for both YCSB workloads, but failed to generate thread-safe or runnable code for the rest.
Figure~\ref{exp:main}(a) ({\bf AI} \greentick) shows the performance of AI-generated B+-tree. Interestingly, AI can generate an optimized B+-tree whose performance matches that of DEX and \name on the YCSB uniform, read-intensive workload. 
However, its performance still falls short of that of hand-crafted DEX and the version optimized by \name on YCSB skew read-intensive workload.

We must admit that closing this gap is just a matter of time or of using a better model. 
Yet, our experience suggests that the central challenge of placing a coding agent in an optimization loop is not merely performance.
Rather, it is the human effort required to write specifications robust enough to prevent reward hacking and, even more critically, to verify the correctness of the generated concurrent code.
For example, during the study, SkyDiscover returned hash table and skip list code that appeared remarkably efficient and scalable at first glance.
Careful manual inspection, however, revealed that they were unsafe, forcing us to discard them
(Figure~\ref{exp:main} {\bf AI} \includegraphics[height=8px]{figures/skull_1f480.png}). 
In other words, using an AI agent shifted work rather than eliminating it: 
developers still had to refine the specification, fix the validator and verify the code.

\name can therefore complement AI-driven approaches. 
An automated research system such as Jitskit may discover a new index structure or algorithm. 
\name can then optimize that structure for disaggregated memory deployment independently. 
This division of labor lets automated research focus on index semantics and algorithmic innovation, while \name handles the recurring, architecture-specific interaction between an index and the memory subsystem. 
As a result, future optimization principles can be reused across indexes rather than re-invented  repeatedly in separate monolithic implementations.

\bibliographystyle{ACM-Reference-Format}
\bibliography{citations}

@inproceedings{ziegler2022scalestore,
  title={ScaleStore: A fast and cost-efficient storage engine using DRAM, NVMe, and RDMA},
  author={Ziegler, Tobias and Binnig, Carsten and Leis, Viktor},
  booktitle={Proceedings of the 2022 International Conference on Management of Data},
  pages={685--699},
  year={2022}
}

@inproceedings{Masstree,
  title={Cache craftiness for fast multicore key-value storage},
  author={Mao, Yandong and Kohler, Eddie and Morris, Robert Tappan},
  booktitle={Proceedings of the 7th ACM european conference on Computer Systems},
  pages={183--196},
  year={2012}
}

@article{Tabular,
  author       = {Ziyi Yan and
                  Mohamed Farouk Drira and
                  Tianxun Hu and
                  Tianzheng Wang},
  title        = {Tabular: Efficiently Building Efficient Indexes},
  journal      = {Proc. {VLDB} Endow.},
  volume       = {18},
  number       = {6},
  pages        = {1991--2004},
  year         = {2025},
  url          = {https://www.vldb.org/pvldb/vol18/p1991-yan.pdf},
  doi          = {10.14778/3725688.3725721},
}

@inproceedings{DMS,
  author       = {Jer{\'{o}}nimo Castrill{\'{o}}n and
                  Jana Giceva and
                  Yu Hua and
                  Kimberly Keeton and
                  Akhil Shekar and
                  Kevin Skadron and
                  Tianzheng Wang and
                  Huanchen Zhang},
  title        = {Declarative Memory Services},
  booktitle    = {16th Conference on Innovative Data Systems Research, {CIDR} 2026,
                  Chaminade, CA, USA, January 18-21, 2026},
  publisher    = {www.cidrdb.org},
  year         = {2026},
  url          = {https://vldb.org/cidrdb/2026/declarative-memory-services.html},
  bibsource    = {dblp computer science bibliography, https://dblp.org}
}

@article{cha2026twosided,
  title         = {Two-sided {RDMA} Striking Back for Disaggregated Memory Databases},
  author        = {Cha, Hokeun and Akella, Aditya and Yu, Xiangyao},
  journal       = {arXiv preprint arXiv:2607.26227},
  year          = {2026},
}

@inproceedings{skydiscover,
  author    = {Shu Liu and Mert Cemri and Shubham Agarwal and
               Alexander Krentsel and Ashwin Naren and Qiuyang Mang},
  title     = {SkyDiscover: A Flexible, Adaptive Framework for {AI}-Driven
               Scientific and Algorithmic Discovery},
  booktitle = {Proceedings of the ACM Conference on AI and Agentic Systems
               (CAIS '26)},
  year      = {2026},
}

@inproceedings{walton1991taxonomy,
  author    = {Walton, Christopher B. and Dale, Alfred G. and Jenevein, Roy M.},
  title     = {A Taxonomy and Performance Model of Data Skew Effects in Parallel Joins},
  booktitle = {Proceedings of the 17th International Conference on Very Large Data Bases (VLDB '91)},
  year      = {1991},
}

@inproceedings{pavlo2012skew,
  author    = {Pavlo, Andrew and Curino, Carlo and Zdonik, Stanley},
  title     = {Skew-Aware Automatic Database Partitioning in Shared-Nothing, Parallel {OLTP} Systems},
  booktitle = {Proceedings of the 2012 International Conference on Management of Data (SIGMOD '12)},
  year      = {2012},
}

@inproceedings{gebara2020stateful,
  author    = {Nadeen Gebara and Alberto Lerner and Mingran Yang and Minlan Yu and Paolo Costa and Manya Ghobadi},
  title     = {Challenging the Stateless Quo of Programmable Switches},
  booktitle = {Proceedings of the 19th ACM Workshop on Hot Topics in Networks (HotNets~'20)},
  year      = {2020},
}

@inproceedings{zhao2026thash,
  author    = {Han Zhao and Chao Wang and Peiqi Yin and Hua Fan and Wenchao Zhou and James Cheng and Eric Lo and Ming-Chang Yang},
  title     = {{THash}: A High-Performance Tiered Hashing for {CXL}-Based Tiered Memory Systems},
  year      = {2026},
}

@online{penguinsolutions2026cxl,
  author = {Penguin Solutions},
  title = {Why AI Needs Compute Express Link (CXL)},
  year = {2026},
  note = {Accessed: 2026-08-18}
}

@online{introl2026cxl,
  author = {Introl},
  title = {CXL 4.0 Infrastructure Planning Guide},
  year = {2026},
  note = {Accessed: 2026-08-18}
}

@online{astoralabs2026azure,
  author = {Astera Labs},
  title = {Astera Labs' Leo CXL Smart Memory Controllers on Microsoft Azure M-series Virtual Machines},
  year = {2026},
  note = {Accessed: 2026-08-18}
}

@online{damnang2026copper,
  author = {{Damnang Research}},
  title = {Copper Has Limits. CXL Has Ambitions. Optics Has Answers.},
  year = {2026},
  url = {https://www.damnang.com/p/copper-has-limits-cxl-has-ambitions},
  note = {Accessed: 2026-08-18}
}

@inproceedings{wang2022sherman,
  author    = {Wang, Qing and Lu, Youyou and Shu, Jiwu},
  title     = {Sherman: A Write-Optimized Distributed {B+Tree} Index on Disaggregated Memory},
  booktitle = {Proceedings of the 2022 International Conference on Management of Data},
  series    = {SIGMOD '22},
  year      = {2022},
}

@article{min2024sephash,
  author    = {Min, Xinhao and Lu, Kai and Liu, Pengyu and Wan, Jiguang and Xie, Changsheng and Wang, Daohui and Yao, Ting and Wu, Huatao},
  title     = {{SepHash}: A Write-Optimized Hash Index on Disaggregated Memory via Separate Segment Structure},
  journal   = {Proceedings of the VLDB Endowment},
  volume    = {17},
  number    = {5},
  year      = {2024},
}

@article{zha2025shard,
  author    = {Zha, Hao and others},
  title     = {Shard: A Scalable and Resize-Optimized Hash Index on Disaggregated Memory},
  journal   = {Proceedings of the VLDB Endowment},
  volume    = {19},
  number    = {4},
  year      = {2025},
}

@article{zhao2026sidle,
  author    = {Zhao, Haoru and Dong, Mingkai and Wu, Fangnuo and Chen, Haibo},
  title     = {{SIDLE}: Tree-Structure Aware Indexes for {CXL}-Based Heterogeneous Memory},
  journal   = {Proceedings of the VLDB Endowment},
  volume    = {19},
  number    = {7},
  year      = {2026},
}

@article{wang2025deft,
  author    = {Wang, Jing and Wang, Qing and Zhang, Yuhao and Shu, Jiwu},
  title     = {Deft: A Scalable Tree Index for Disaggregated Memory},
  journal   = {Proceedings of the 20th ACM European Conference on Computer Systems},
  series    = {EuroSys '25},
  year      = {2025},
}

@article{wang2023case,
  author    = {Wang, Ruihong and Wang, Jianguo and Idreos, Stratos and {\"O}zsu, M. Tamer and Aref, Walid G.},
  title     = {The Case for Distributed Shared-Memory Databases with {RDMA}-Enabled Memory Disaggregation},
  journal   = {Proceedings of the VLDB Endowment},
  volume    = {16},
  number    = {1},
  year      = {2022},
}

@article{almaruf2023survey,
  author = {Al Maruf, Hasan and Chowdhury, Mosharaf}, 
  title = {Memory Disaggregation: Advances and Open Challenges}, 
  year = {2023}, 
  volume = {57}, 
  number = {1},
  journal = {SIGOPS Oper. Syst. Rev.}
}

@article{sjtu2025survey,
  author = {Wang, Jing and Li, Chao and Wang, Taolei and Guo, Jinyang and Yang, Hanzhang and Zhuansun, Yiming and Guo, Minyi},
  title = {Survey of Disaggregated Memory: Cross-Layer Technique Insights for Next-Generation Datacenters},
  year = {2026},
  issue_date = {September 2026},
  volume = {58},
  number = {12},
  issn = {0360-0300},
  journal = {ACM Comput. Surv.}
}

@inproceedings{maruf2023tpp,
  author    = {Al Maruf, Hasan and Wang, Yizhou and Stuedi, Patrick and Zhang, Yiying},
  title     = {{TPP}: Transparent Page Placement for {CXL}-Enabled Tiered-Memory},
  booktitle = {Proceedings of the 28th ACM International Conference on Architectural Support for Programming Languages and Operating Systems},
  series    = {ASPLOS '23},
  volume    = {3},
  year      = {2023},
}

@article{zhong2024unimem,
  author    = {Zhong, Yizhou and others},
  title     = {{UniMem}: Redesigning Disaggregated Memory within A Unified Memory Abstraction},
  journal   = {Proceedings of the 2024 USENIX Annual Technical Conference},
  year      = {2024},
}

@article{clove2026,
  author    = {Sam Son and Zhihong Luo and Wen Zhang and Sylvia Ratnasamy and Scott Shenker},
  title     = {{Clove}: Object-Level {CXL} Memory Management in Managed Runtimes},
  journal   = {arXiv preprint arXiv:2605.20370},
  year      = {2026}
}

@inproceedings{li2025beehive,
  author    = {Li, Quanxi and Huang, Hong and Liu, Ying and Xia, Yanwen and Zhang, Jie and Zhou, Mosong and Feng, Xiaobing and Cui, Huimin and Chen, Quan and Shan, Yizhou and Wang, Chenxi},
  title     = {{Beehive}: A Scalable Disaggregated Memory Runtime Exploiting Asynchrony of Multithreaded Programs},
  booktitle = {Proceedings of the 22nd USENIX Symposium on Networked Systems Design and Implementation},
  series    = {NSDI '25},
  year      = {2025},
}

@inproceedings{calciu2021kona,
  author    = {Calciu, Irina and others},
  title     = {Rethinking Software Runtimes for Disaggregated Memory},
  booktitle = {Proceedings of the 26th ACM International Conference on Architectural Support for Programming Languages and Operating Systems},
  series    = {ASPLOS '21},
  year      = {2021},
}

@article{calciu2023kona,
  author    = {Calciu, Irina and others},
  title     = {Using Local Cache Coherence for Disaggregated Memory},
  journal   = {SIGOPS Oper. Syst. Rev.},
  year      = {2023},
}

@article{pimdal2025,
  title={PIMDAL: Mitigating the Memory Bottleneck in Data Analytics using a Real Processing-in-Memory System},
  author={Manos Frouzakis and Juan Gómez-Luna and Geraldo F. Oliveira and Mohammad Sadrosadati and Onur Mutlu},
  year={2025},
  eprint={2504.01948},
  archivePrefix={arXiv},
  primaryClass={cs.AR},
}

@inproceedings{bernhardt2023pimdb,
  author = {Bernhardt, Arthur and Koch, Andreas and Petrov, Ilia},
  title = {pimDB: From Main-Memory DBMS to Processing-In-Memory DBMS-Engines on Intelligent Memories},
  year = {2023},
  booktitle = {Proceedings of the 19th International Workshop on Data Management on New Hardware},
  numpages = {9},
  location = {Seattle, WA, USA},
  series = {DaMoN '23}
}

@misc{pimoverview2025,
  author    = {Emergent Mind},
  title     = {Processing-in-Memory ({PIM}) Overview},
  howpublished = {\url{https://www.emergentmind.com/topics/processing-in-memory-pim}},
  year      = {2025},
  note      = {Accessed: 2026-08-11}
}

@inproceedings{ruan2020aifm,
  author    = {Ruan, Zhen and others},
  title     = {{AIFM}: High-Performance, Application-Integrated Far Memory},
  booktitle = {Proceedings of the 14th USENIX Symposium on Operating Systems Design and Implementation},
  series    = {OSDI '20},
  year      = {2020},
}

@mastersthesis{he2023thesis,
  author       = {He, Zijian},
  title        = {Towards a Transparent and Efficient Far Memory System},
  school       = {University of California, San Diego},
  year        = {2023},
  type        = {Master's thesis},
  note        = {Advisor: Yiying Zhang}
}

@inproceedings{tauro2024trackfm,
  author    = {Tauro, Brian R. and Suchy, Brian and Gao, Simone and Dinda, Peter A.},
  title     = {{TrackFM}: Far-out Compiler Support for a Far Memory World},
  booktitle = {Proceedings of the 29th ACM International Conference on Architectural Support for Programming Languages and Operating Systems},
  series    = {ASPLOS '24},
  volume    = {1},
  year      = {2024},
}

@article{m5_2025,
  author    = {Sun, Yan and Kim, Jongyul and Yu, Zeduo and Zhang, Jiyuan and Chai, Siyuan and Kim, Michael Jaemin and Nam, Hwayong and Park, Jaehyun and Na, Eojin and Yuan, Yifan and Wang, Ren and Ahn, Jung Ho and Xu, Tianyin and Kim, Nam Sung},
  title     = {{M5}: Mastering Page Migration and Memory Management for {CXL}-based Tiered Memory Systems},
  journal   = {Proceedings of the 30th ACM International Conference on Architectural Support for Programming Languages and Operating Systems},
  series    = {ASPLOS '25},
  year      = {2025}
}

@article{flexmem2024,
  author    = {Xu, Dong and Ryu, Junhee and Baek, Jinho and Shin, Kwangsik and Su, Pengfei and Li, Dong},
  title     = {FlexMem: Adaptive Page Profiling and Migration for Tiered Memory},
  journal   = {Proceedings of the 2024 USENIX Annual Technical Conference},
  year      = {2024},
}

@article{aceso2024,
  author    = {Hu, Zhisheng and Zuo, Pengfei and Chen, Yizou and Wang, Chao and Hu, Junliang and Yang, Ming-Chang},
  title     = {Aceso: Achieving Efficient Fault Tolerance in Memory-Disaggregated Key-Value Stores},
  journal   = {Proceedings of the 29th ACM Symposium on Operating Systems Principles},
  series    = {SOSP '24},
  year      = {2024},
}

@article{cxlsurvey2025,
  author = {Yonggui Liang and Tao Huang and Kexin Chen and Shubao Yu and Yiwen Sun and Zhengpei Liu and Yunfu Wang and Boyi Dong and Xueguo Yan},
  title = {Innovation in Computational Architecture: Opportunities and Challenges of CXL Memory Disaggregation Technology in Intelligent Computing Centers},
  year = {2026},
  journal = {Tsinghua Science and Technology},
  volume = {31},
  number = {4},
}

@inproceedings{cxlperf2025,
  author={Wang, Xi and Liu, Jie and Wu, Jianbo and Yang, Shuangyan and Ren, Jie and Shankar, Bhanu and Li, Dong},
  booktitle={2025 IEEE International Parallel and Distributed Processing Symposium (IPDPS)},
  title={Performance Characterization of CXL Memory and Its Use Cases},
  year={2025},
}

@misc{cxltype2_2025,
  author    = {CXL Consortium},
  title     = {{CXL} Type 2: Use Cases for Active Memory Tiering and Near-Memory Accelerators},
  howpublished = {\url{https://computeexpresslink.org/blog/cxl-type-2-use-cases-for-active-memory-tiering-and-near-memory-accelerators-3989/}},
  year      = {2025},
  note      = {Accessed: 2026-08-11}
}

@article{lu2024dex,
  author    = {Baotong Lu and Kaisong Huang and Chieh-Jan Mike Liang and
               Tianzheng Wang and Eric Lo},
  title     = {DEX: Scalable Range Indexing on Disaggregated Memory},
  journal   = {Proceedings of the VLDB Endowment},
  volume    = {17},
  number    = {10},
  year      = {2024},
}

@inproceedings{guo2023mira,
  author    = {Zhiyuan Guo and Zijian He and Yiying Zhang},
  title     = {Mira: A Program-Behavior-Guided Far Memory System},
  booktitle = {Proceedings of the 29th Symposium on Operating Systems Principles},
  series    = {SOSP '23},
  year      = {2023},
}

@article{lu2025chash,
  author    = {Mengting Lu and Gaocong Liu and others},
  title     = {CHash: A High Cost-Performance Hash Design for CXL-Based Disaggregated Memory System},
  journal   = {Proceedings of the ACM on Measurement and Analysis of Computing Systems},
  year      = {2025},
}

@inproceedings{shan2018legoos,
  author    = {Yizhou Shan and Yutong Huang and Yilun Chen and Yiying Zhang},
  title     = {{LegoOS}: A Disseminated, Distributed {OS} for Hardware Resource Disaggregation},
  booktitle = {13th USENIX Symposium on Operating Systems Design and Implementation (OSDI 18)},
  year      = {2018},
  month     = {October},
}

@misc{cheng2025letbarbariansinai,
  title={Let the Barbarians In: How AI Can Accelerate Systems Performance Research},
  author={Audrey Cheng and Shu Liu and Melissa Pan and Zhifei Li and Shubham Agarwal and Mert Cemri and Bowen Wang and Alexander Krentsel and Tian Xia and Jongseok Park and Shuo Yang and Jeff Chen and Lakshya Agrawal and Ashwin Naren and Shulu Li and Ruiying Ma and Aditya Desai and Jiarong Xing and Koushik Sen and Matei Zaharia and Ion Stoica},
  year={2025},
  eprint={2512.14806},
  archivePrefix={arXiv},
  primaryClass={cs.SE},
}

@misc{Belcic_Stryker_2026,
  title={What is Loop Engineering?},
  journal={IBM},
  author={Belcic, Ivan and Stryker, Cole},
  year={2026},
  month={Jul}
}

@misc{wehrstein2026bespokeolap,
  title={Bespoke OLAP: Synthesizing Workload-Specific One-size-fits-one Database Engines},
  author={Johannes Wehrstein and Timo Eckmann and Matthias Jasny and Carsten Binnig},
  year={2026},
  eprint={2603.02001},
  archivePrefix={arXiv},
  primaryClass={cs.DB},
}

@misc{liu2026timejits,
  title={The Time is Here for Just-in-Time Systems: Challenges and Opportunities},
  author={Shu Liu and Alexander Krentsel and Shubham Agarwal and Mert Cemri and Ziming Mao and Soujanya Ponnapalli and Alexandros G. Dimakis and Sylvia Ratnasamy and Matei Zaharia and Aditya Parameswaran and Ion Stoica},
  year={2026},
  eprint={2605.24096},
  archivePrefix={arXiv},
  primaryClass={cs.DB},
}

@misc{zhong2025impossiblebenchmeasuringllmspropensity,
  title={ImpossibleBench: Measuring LLMs' Propensity of Exploiting Test Cases},
  author={Ziqian Zhong and Aditi Raghunathan and Nicholas Carlini},
  year={2025},
  eprint={2510.20270},
  archivePrefix={arXiv},
  primaryClass={cs.LG},
}

@online{pmemio2023cxlecosystem,
  author       = {{pmem.io}},
  title        = {Exploring the Software Ecosystem for Compute Express Link (CXL) Memory},
  year         = {2023},
  month        = may,
  day          = {25},
  urldate      = {2026-08-18},
  organization = {pmem.io},
  note         = {Blog post}
}

@techreport{lenovo2025cxlimplementation,
  author      = {{Lenovo}},
  title       = {Implementing CXL Memory on Linux on ThinkSystem V4 Servers},
  institution = {Lenovo Press},
  year        = {2025},
  month       = mar,
  day         = {21},
  number      = {LP2184},
  urldate     = {2026-08-18},
  note        = {23 pages}
}

@online{shilov2026metacxl,
  author  = {Shilov, Anton},
  title   = {Meta Fights Soaring Hardware Costs by Reusing Old DDR4 Server Memory in New DDR5-only Servers --- Custom CXL 2.0 Chip Marries Legacy DDR4-2400 with Cutting-Edge DDR5-6400},
  year    = {2026},
  month   = jun,
  day     = {30},
  urldate = {2026-08-18},
  organization = {Tom's Hardware}
}

@article{wang2024cache,
  author  = {Ruihong Wang and Jianguo Wang and Walid G. Aref},
  title   = {Cache Coherence Over Disaggregated Memory},
  journal = {Proceedings of the VLDB Endowment},
  volume  = {18},
  number  = {9},
  year    = {2025},
}

@inproceedings{leanstore,
  title={Leanstore: In-memory data management beyond main memory},
  author={Leis, Viktor and Haubenschild, Michael and Kemper, Alfons and Neumann, Thomas},
  booktitle={2018 IEEE 34th International Conference on Data Engineering (ICDE)},
  year={2018},
  organization={IEEE}
}

@article{hart2007performance,
  title={Performance of memory reclamation for lockless synchronization},
  author={Hart, Thomas E and McKenney, Paul E and Brown, Angela Demke and Walpole, Jonathan},
  journal={Journal of Parallel and Distributed Computing},
  volume={67},
  number={12},
  year={2007},
}

@inproceedings{cadar2008klee,
  author    = {Cadar, Cristian and Dunbar, Daniel and Engler, Dawson R.},
  title     = {KLEE: Unassisted and Automatic Generation of High-Coverage Tests for Complex Systems Programs},
  booktitle = {Proceedings of the 8th USENIX Symposium on Operating Systems Design and Implementation (OSDI)},
  year      = {2008},
}

\end{document}